\documentclass[sigconf, nonacm]{acmart}

\def\BibTeX{{\rm B\kern-.05em{\sc i\kern-.025em b}\kern-.08emT\kern-.1667em\lower.7ex\hbox{E}\kern-.125emX}}
    
\usepackage{multirow}

\begin{document}

%
\title[Application Interference on Dragonfly+]{Modeling and Analysis of Application Interference on Dragonfly+
}
\thanks{The paper was accepted by ACM SIGSIM-PADS'19. If you use this work, please cite the paper as: Kang Y, Wang X, McGlohon N, Mubarak M, Chunduri S, Lan Z. Modeling and analysis of application interference on dragonfly+. In Proceedings of the 2019 ACM SIGSIM Conference on Principles of Advanced Discrete Simulation 2019 May 29 (pp. 161-172).}

%
\author{Yao Kang}
\affiliation{%
  \institution{Illinois Institute of Technology}
  \streetaddress{}
  \city{Chicago, IL}
 \country{USA}
}
\email{ykang17@hawk.iit.edu}

\author{Xin Wang}
\affiliation{%
  \institution{Illinois Institute of Technology}
  \streetaddress{}
  \city{Chicago, IL}
  \country{USA}
  }
\email{xwang149@hawk.iit.edu}

\author{Neil McGlohon}
\affiliation{%
  \institution{Rensselaer Polytechnic Institute}
  \city{Troy, NY}
  \country{USA}
}
\email{mcglon@rpi.edu}

\author{Misbah Mubarak}
\affiliation{%
 \institution{Argonne National Laboratory}
 \streetaddress{}
 \city{Lemont, IL}
 \country{USA}
}
\email{mmubarak@anl.gov}
 
\author{Sudheer Chunduri}
\affiliation{%
 \institution{Argonne National Laboratory}
 \streetaddress{}
 \city{Lemont, IL}
 \state{}
 \country{USA}
 }
 \email{sudheer@anl.gov}
 
\author{Zhiling Lan}
\affiliation{%
  \institution{Illinois Institute of Technology}
  \streetaddress{}
  \city{Chicago, IL}
  \country{USA}
  }
\email{lan@iit.edu}

%

%
\begin{abstract}

Dragonfly class of networks are considered as promising interconnects for next-generation supercomputers.
While Dragonfly+ networks offer more path diversity than the original Dragonfly design, they are still prone to performance variability due to their hierarchical architecture and resource sharing design.
Event-driven network simulators are indispensable tools for navigating complex system design. In this study, we quantitatively evaluate a variety of application communication interactions on a 3,456-node Dragonfly+ system by using the CODES toolkit.
This study looks at the impact of communication interference from a user's perspective.
Specifically, for a given application submitted by a user, we examine how this application will behave with the existing workload running in the system under different job placement policies.
Our simulation study considers hundreds of experiment configurations including four target applications with representative communication patterns under a variety of network traffic conditions.
Our study shows that intra-job interference can cause severe performance degradation for communication-intensive applications. 
Inter-job interference can generally be reduced for applications with one-to-one or one-to-many communication patterns through job isolation. Application with one-to-all communication pattern is resilient to network interference.

\end{abstract}

%
%


%

%
\maketitle

\section{Introduction}

Scientific simulations used to take us years to finish can now be completed in seconds on supercomputers or high-performance computing (HPC) systems. According to the recent Top500 list\cite{top500}, the fastest supercomputer is capable of 200 PFlop/s computing rate, and all of the top 4 systems have a peak rate greater than 100 PFlop/s.
In order to achieve petascale or higher performance, 
state-of-the-art supercomputers are deployed with millions of cores and thousands of nodes. Interconnect network plays a critical role in these systems as it serves as a "central nervous system" for data exchange among system resources. In the past years, the ever-increasing need for higher bandwidth, lower latency, and higher message rate has driven the deployment of Dragonfly networks.

The conception of Dragonfly topology was introduced in 2008\cite{df}. 
It is an hierarchical network topology dividing a system's compute nodes into several identical groups. Global links are used to connect all groups in an all-to-all manner.
In \cite{df}, the network performance analysis is based on an all-to-all, one-dimensional intra-group connection. Therefore, this network architecture is often referred as 1D-Dragonfly.
The all-to-all intra-group connection somehow limits the maximum number of compute nodes a group can hold and hence limits the maximum system size. 

The Cray Cascade system\cite{cascade} is a distributed memory system based on the Dragonfly network topology. Instead of using 1-dimen-\\sional all-to-all intra-group connection, the Cray Cascade system deploys a two-dimensional, partially all-to-all intra-group connection. 
Within a group, routers are arranged into multiple rows and columns. The routers on the same rows or the same columns are all-to-all connected and there is no direct connection between routers that do not share a common row or column. The two dimensions in a group (i.e. row dimension and column dimension) makes this network commonly be referred as 2D-Dragonfly. 
Compared with the conventional 1D-Dragonfly, 2D-Dragonfly can support a larger system size because it removes the connections between the routers not on the same dimension and uses the saved ports for additional compute nodes. However, this advantage comes at the cost of increasing network diameter and making 2D-Dragonfly a 5-hop topology.

Dragonfly+\cite{dfp} or Megafly\cite{megafly} is a new Dragonfly variant proposed to support a large system size while achieving a low network diameter. 
Dragonfly+ adopts a complete bipartite-graph as its intra-group connection. Similar to 1D-Dragonfly, it is a diameter-three topology. 
Additionally, using the same hardware, Dragonfly+ can support 
a system size four times larger than 1D-Dragonfly.
This means that under the same system size, path diversity on Dragonfly+ is augmented by increasing the number of minimal paths between compute nodes. 
The low diameter and large system size characteristics make Dragonfly+
a candidate for exascale interconnect topology\cite{exaconnect}.
The Niagara system\cite{niagara} at SciNet supercomputing centre is a production system adopting Dragonfly+ topology.

Network resources in Dragonfly systems are shared among user applications in a multi-user computing environment. Network interference among applications for shared network resources can cause severe variance in message arrival times due to ephemeral contention events\cite{runtorun}.
In general, the network interference can be classified into two categories: {\it intra-job interference} and {\it inter-job interference}. 
Intra-job interference, also referred as self-congestion, denotes the competition for the shared network resources among parallel processes belonging to the same job, and inter-job interference means the network resources competition between parallel processes from different jobs.
Previous studies have unveiled that performance variation is a serious issue on 1D-Dragonfly and 2D-Dragonfly systems\cite{xin}\cite{bully}. Nevertheless, {\it little work has been done for Dragonfly+ systems} and this study is intended to bridge the gap. 

In this study, we analyze intra-job and inter-job communication interference on Dragonfly+ by using the CODES network simulator\cite{codes}. 
Specifically, our analysis emphasizes the interference from a user's perspective. 
When a user submits his/her application for execution on a production system, we analyze how this user job (denoted as \textit{target application}) behaves under the existing running applications (denoted as \textit{background application}).
Toward this end, we enhance the current CODES Dragonfly+ module by implementing the Fully Progressive Adaptive Routing adopted in the production Dragonfly+ system, and adding a mechanism for simulating multiple synthetic workloads.
The enhanced CODES version enables us to simulate a 3,456-node Dragonfly+ system under a variety of background traffics and target applications. We examine four target applications, each with a distinct communication pattern. For each target application, we consider nine combinations of message frequencies and message sizes, ranging from small-sized messages in a low frequency to large-sized messages in a high frequency. Moreover, we consider two job placement policies for allocating the target application onto the system with various background traffic conditions, one isolating the application from the background application and the other mixing the target application with the background application. We examine target application in terms of {\it message latency} (defined as message traveling time from its source to the destination) under various computing environments. 

We make several key findings from the extensive simulation study: 
First, communication-intensive applications are severely affected by intra-job interference, especially when they transmit messages at a rate greater than the available network bandwidth.
Intra-job interference can be mitigated with contiguous job placement for applications with 3D stencil pattern or with random placement for applications with tornado pattern.
Second, for the applications with one-to-one or one-to-many communication pattern, performance variation caused by inter-job interference can generally be mitigated through group-level workload isolation such that different applications do not share the same Dragonfly+ groups. 
Third, application with one-to-all communication pattern such as broadcasting is resilient to inter-job interference, and existing network traffic can hardly affect its performance.

The rest of the paper is organized as follows. Section \ref{sec:background} presents the Dragonfly+ topology, the CODES simulator and related work. Section \ref{sec:method} provides CODES Dragonfly+ module enhancement with experimental design and configurations. Simulation result analysis is given in Section \ref{sec:result} before the conclusion in Section \ref{sec:conclude}.

\section{Background and Related Work} \label{sec:background}
In this section, we provide an overview of the Dragonfly+ topology including system architecture, network diameter, and the maximum supported system size. We also introduce the CODES simulator and the 3,456-node Dragonfly+ system simulated in this study. 

\subsection{Dragonfly+ Topology}

Dragonfly+ \cite{dfp}, also referred as Megafly \cite{megafly}, is a new variant of Dragonfly topology. The key difference between these two networks is the connection arrangement within groups. 
Routers within local groups in a 1D-Dragonfly network are connected in an all-to-all manner. Dragonfly+, however, has a local group connection structure that forms a complete bipartite graph. 
Both typologies have their groups all-to-all connected.

\begin{figure}[htbp]
      \centering
      \includegraphics[width=1\linewidth]{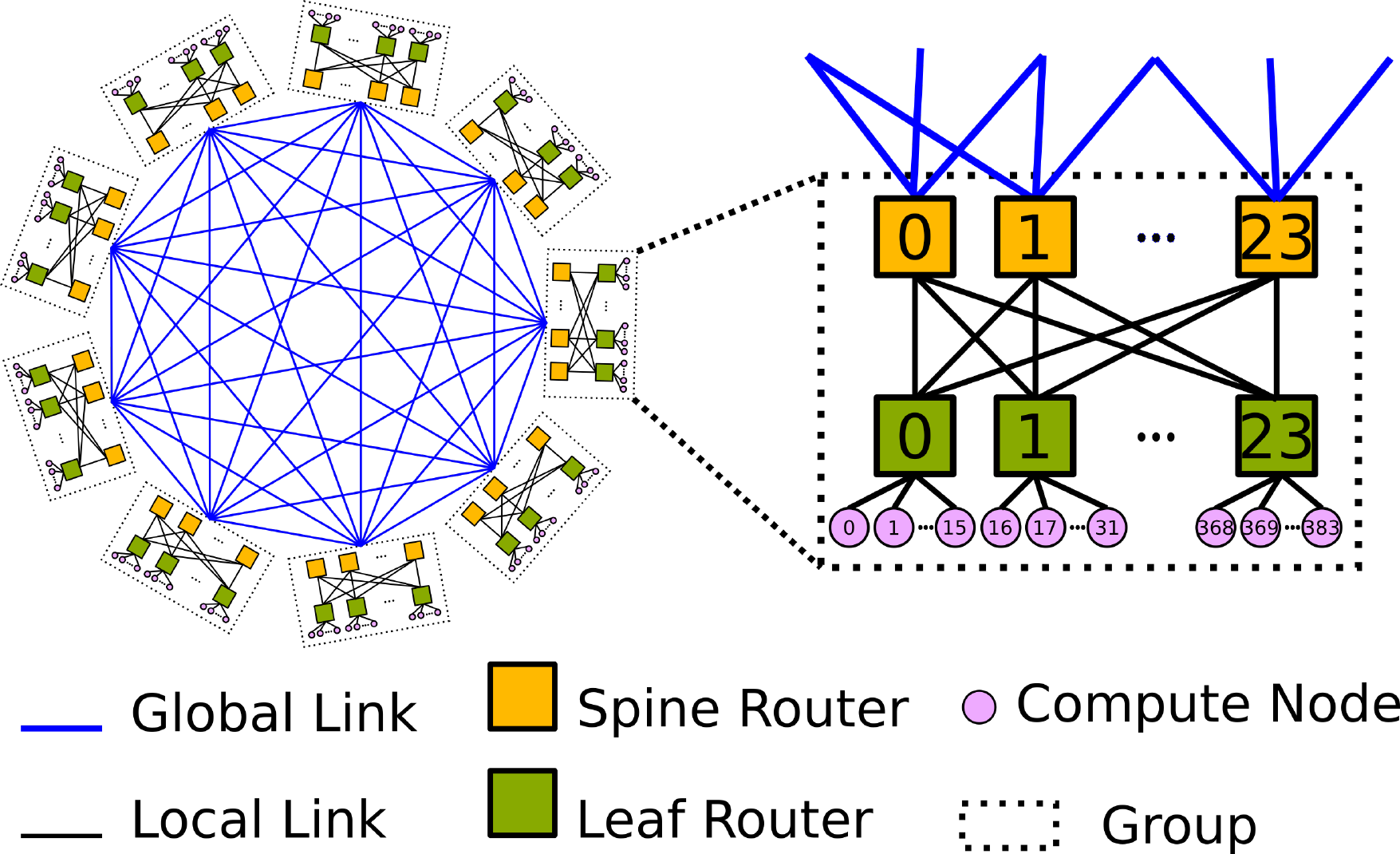}
      \caption{A 3,456-node Dragonfly+ system}
      \label{fig:dfp}
\end{figure}

Figure \ref{fig:dfp} depicts a 3,456-node Dragonfly+ system simulated in this study, along with a detailed bipartite router connection in a group. Groups are connected via global links (in blue) and routers are connected via local links (in black). There are two types of routers in a Dragonfly+ group: leaf routers shown as green boxes and spine routers shown as yellow boxes.
Leaf routers only have local links to spine routers and links to compute nodes, shown as pink circles in Figure \ref{fig:dfp}. 
Spine routers, on the other hand, only have local links to leaf routers and global links to spine routers in other groups.
In a single group, each leaf router is connected to all spine routers and each spine router has connections to all leaf routers. There is no connection between any routers of the same type. This configuration makes Dragonfly+ intra-group connection into a complete bipartite-graph.

Dragonfly+ is a diameter three network topology. It means that any two nodes are connected through 3 links: 2 local links in the source and the destination groups and 1 global link between them. The links between compute nodes and leaf routers are not counted as they are inevitable and independent of network topology.

When using r-radix router, the maximum group size can be achieved by having $r/2$ spine routers and $r/2$ leaf routers such that each leaf router is attached with $r/2$ compute nodes and every spine router has $r/2$ global links. Therefore, Dragonfly+ maximum system size can be determined by the following formula:

\begin{equation} \label{eq:dfp}
         S_{+} = (\frac{r}{2})(\frac{r}{2})*((\frac{r}{2})(\frac{r}{2})+1)
\end{equation}

For 1-D dragonfly network, the maximum group size is achieved  when each router has $r/4$ ports for global link connections, $r/4$ ports for compute nodes and $r/2$ ports for local links. Therefore the maximum system size can be calculated as follows:

\begin{equation} \label{eq:df}
    S_{1D} = \frac{r}{4}(\frac{r}{2}+1)*(\frac{r}{4}(\frac{r}{2}+1)+1)
\end{equation} 

By comparing Equation (\ref{eq:dfp}) and (\ref{eq:df}), Dragonfly+ can support as much as 4x compute nodes than 1D-Dragonfly with the same radix routers. 

\begin{equation}
    \lim_{r\to\infty} \frac{S_{+}}{S_{1D}}=4
\end{equation}

This is a key advantage of Dragonfly+ compared with 1D-Drago-\\nfly: using the same radix routers and connection links, 
Dragonfly+ supports a larger system size. In other words, under the same system size, Dragonfly+ can provide more minimal paths between two compute nodes. As network resources are shared among jobs on system,  an increase in path diversity between compute nodes leads to less network contention between jobs, which eventually causing less job runtime uncertainty and improve overall system performance.

In this study, we simulate a 3,456-node Dragonfly+ system as illustrated in Figure \ref{fig:dfp} with 9 groups using 48-radix routers. Each group has 24 spine routers and 24 leaf routers. There are 16 compute nodes on each leaf router and 16 global links on each spine router. As a result, each spine router has 2 global links connecting to any other group and 48 global links between any two groups. Network bandwidth is configured according to the Cray Cascade system\cite{cascade} with $4.37GiB/s$ global link bandwidth, $5.25GiB/s$ local link bandwidth between routers and $16GiB/s$ channel bandwidth between leaf router and compute nodes.


\subsection{CODES Simulator}

CODES (Enabling \textbf{CO-D}esign of \textbf{E}xascale \textbf{S}torage Systems) is an event-driven network simulator\cite{codes}. 
It provides a set of HPC interconnect models for researchers to simulate different system designs, and ROSS\cite{ross} serves as the underlying event-driven simulation framework for CODES. 

CODES has been validated against real systems or cycle accurate simulator in the past studies\cite{1dfnetworkinterfer}\cite{thetavalid}. Many studies have used CODES for different network topologies analysis  \cite{codes_fattree}\cite{codes_dragonfly}\cite{codes_torus}. 
CODES offers a Dragonfly+ module since the release of version 1.0.0, which is enhanced and used in this study.


\subsection{Related Work}

Simulation study from Jain et al. \cite{dfrandplace} showed that using random job placement policy on Dragonfly system can help spread communication traffic across the network and reduce hot-spots. However, random job placement on Dragonfly system tends to show a higher performance variability issue\cite{1dfnetworkinterfer}.
Yang et al. \cite{bully} studied the "bully" effect on 1D-Dragonfly system, his study showed that a strong overall network performance is achieved by impairing less communication intensive applications.
Chunduri et al. \cite{runtorun} studied network interference on the Cray Cascade production system, and unveiled that user can experience a run-to-run job performance variation in real life.
Wang et al. \cite{xin,wang2020union} studied the Cray Cascade system through simulation, and demonstrated that intra-job interference can be mitigated with contiguous job placement for low message load applications.

Dragonfly+ topology is introduced by Shpiner et al. \cite{dfp}. as a high performance interconnect network. 
Flajslik et al. \cite{megafly} compared Dragonfly+ with 1D-Dragonfly system showing that Dragonfly+ can provide a higher path diversity and a better throughput. Their study also discussed system cost and power design problem with global and local tapering options on the Dragonfly+ system.

Although a lot of studies have been conducted for network interference on Dragonfly systems, such problem is not well investigated on Dragonfly+ topology, which motivates this study.

\section{Methodology} \label{sec:method}

The goal of this study is to find out how a user job can be impacted by other jobs sharing the Dragonfly+ system under different conditions.
Hence the performance of the user job (i.e. \textit{target application}) is extensively examined in our experiments. 
In this study, we analyze four types of target applications, each 
with a distinct communication pattern. For each target application, we consider a variety of message sizes and message frequencies. 
We use a background application to simulate network traffic generated by other jobs (other than the target application). The background application generates MPI messages under a variety of communication intensities. In this section, we present the CODES Dragonfly+ module enhancement, the background application, and the target applications. 


\subsection{Dragonfly+ Module Enhancement}

The current CODES release provides the Dragonfly+ network model. For the purpose of the intra- and inter-job interference study, we make several changes to the CODES simulator, which are described below.

\subsubsection{\textbf{Balancing global link connection}}

Neither Dragonfly+\cite{dfp} nor Megafly\cite{megafly} paper fully discussed global link connections between groups. The only specification is that Dragonfly+ topology requires the same number of global links between any two groups, which is 48 in this study.

The CODES dragonfly+ module supports arbitrary inter-group connection arrangement by taking an external configuration file. This configuration file can be generated with a provided script. When a spine router has multiple global links connecting to another group, the current script generates inter-group connection by having all available global links connected to one spine router in the destination group. 
When a large quantity of data is transmitted between two groups, this repeated connection between two spine routers in source and destination group can potentially harm network performance by reaching the receiver router's maximum routing capacity and congesting local link when all transmitted data has the same destination.

We change the global link connection into a linear assignment way as shown in Figure 2 for a more balanced inter-group connection. 

\begin{figure}[ht]
	\centering
	\includegraphics[width=1\linewidth]{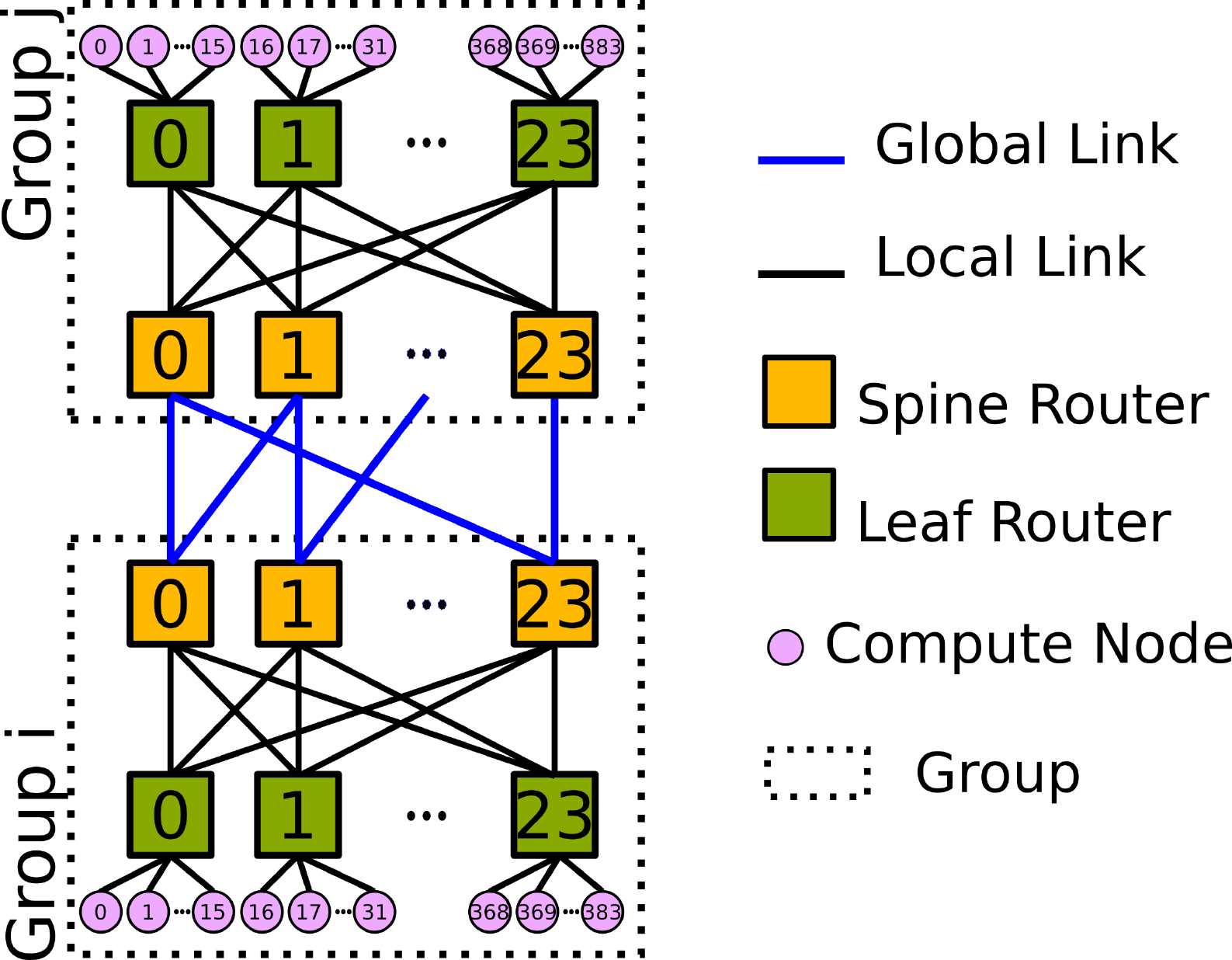}
	\caption{Global link connection between group $i$ and group $j$, where $ 0\leq i < j \leq 8$.}
	\label{fig:gcc}
\end{figure}

Figure \ref{fig:gcc} illustrates the global link connections between group $i$ and $j$, where $ 0\leq i < j \leq 8$. Router $k$ in group $i$ is connected to router $k'$ in group $j$ with equation:
$k'=(k+l)\bmod num\_spine$
, where $l$ is 0 or 1 and $num\_spine$ is the number of spine routers in a group, which is a half of router radix for a symmetric bipartite intra-group architecture.

\subsubsection{\textbf{Implementing Fully Progressive Adaptive Routing}} \text{ }
Dragonfly+ relies on Fully Progressive Adaptive Routing (FPAR)\cite{dfp} to balance traffic loads among network links. The essence of adaptive routing is to let router choose between minimal and non-minimal paths depending on their congestion condition with the preference of lower hops routing path. 

\begin{figure}[htbp]
      \centering
      \includegraphics[width=1\linewidth]{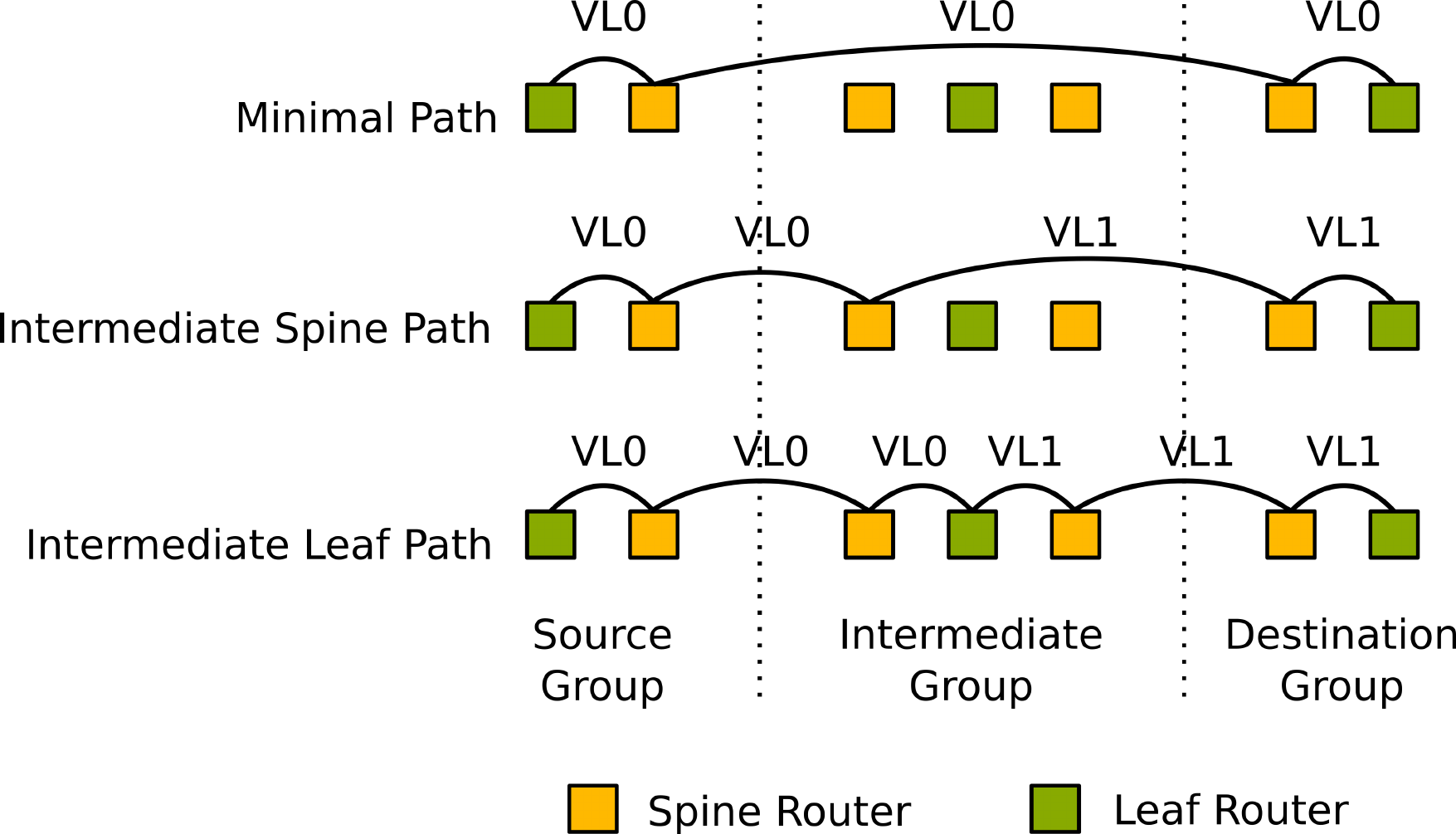}
      \caption{Fully Progressive Adaptive Routing Paths}
      \label{fig:fpar}
\end{figure}

As shown in Figure \ref{fig:fpar}, FPAR lets router choose between three paths for message forwarding: one minimal path, and two non-minimal paths (intermediate spine/leaf path). Minimal path is the shortest routing path by forwarding messages from the source group directly to the destination group through three links, whereas intermediate leaf path is the longest route with message transferred through three routers in an intermediate group. 
Figure \ref{fig:fpar} also depicts how FPAR implements deadlock-free routing with the help of two Virtual Lanes (VL) or Virtual Channels (VC). 
FPAR only uses VL0 on minimal path, and VL0 and VL1 on non-minimal paths. VL is only switched from 0 to 1 on non-minimal path by either an intermediate spine router or an intermediate leaf router once FPAR decides to send the message minimally to the destination node. Router receives a message from VL1 will only forward it minimally without considering any non-minimal paths. This mechanism avoids cyclic routing path in the intermediate group. 
FPAR makes routing decision between three paths (one minimal and two non-minimal) by firstly checking the available paths and identify its corresponding port number on the current router. In the case of the message is from VL1, only minimal paths are considered. Next a score as the port queue occupancy normalized by the queue length is attached to each available path. Final decision is made by comparing port score with a predefined threshold \textit{T}. A longer path with score smaller than \textit{T} will be chosen if and only if all shorter paths' scores are greater than \textit{T}. When all shorter and longer paths are occupied with a score greater than \textit{T}, the smallest score shorter path will be chosen.

The CODES Dragonfly+ network model comes with a default progressive adaptive routing algorithm based on the comparison between absolute channel occupancy values. It provides similar but not exactly identical behavior as FPAR. Based on this routing function, we implement FPAR to make the simulator have the same routing behavior as described in \cite{dfp}. In this study, the predefined threshold \textit{T} is set to $50\%$ as non-minimal paths require twice more global links than minimal path.

\subsubsection{\textbf{Supporting Multiple Synthetic Workloads}}
CODES currently supports the simulation of one synthetic workload occupying the entire system. 
We enhance CODES to make it support multiple synthetic workloads so as to better emulate a production computing environment.

Our modification makes the CODES synthetic simulator read two additional configuration files: workload file and allocation file. Different jobs are defined in the workload file, one line per job, with their detailed information such as MPI traffic pattern, job size (number of processes), message interval time, message size, and number of messages to be sent. The allocation file takes care of job placement. In our experiments, we only place one MPI process per node. The allocation file provides information such as which MPI process should be placed on which node.

In this study we focus on two simultaneous workloads: one is the target application, and the other is the background application generating MPI traffics at different intensities. 

\newpage


\subsection{Theoretical Global Link Load (TGLL)}

In order to have a comprehensive network interference study, parameters related to MPI messages should be taken into consideration. Throughout this study, we have MPI message size, message interval, and global link bandwidth as experimental parameters.
Message interval denotes the time duration between two consecutive messages' issuing time. A shorter message interval means that the application injects traffic into the network more frequently and gives a higher communication intensity.
In order to represent different network usage scenarios, we define a variable called Theoretical Global Link Load (TGLL):

\begin{equation} \label{eq:tgll}
    TGLL = \frac{Msg\_size/Msg\_interval}{Global\_Link\_Bandwidth}
\end{equation}

TGLL indicates communication intensity with respect to the usage of global link bandwidth and provides an intuitive indication of global link load. 

In this study, we use TGLL to quantify communication intensity for both target application and background application under three cases: \textit{underutilized (TGLL $20\% - 50\%$), near-saturated (TGLL $70\% - 90\%$) and overloaded (TGLL $>100\%$)}.

\subsection{Background Application}

The background application is used to generate network traffic as a mix of MPI communications from all jobs other than the target application. Background messages are generated with the source and destination pairs in the first three groups (Group 0, 1, 2 ) at different intensities. 

\begin{figure}[htbp]
	\centering
	\includegraphics[width=1\linewidth]{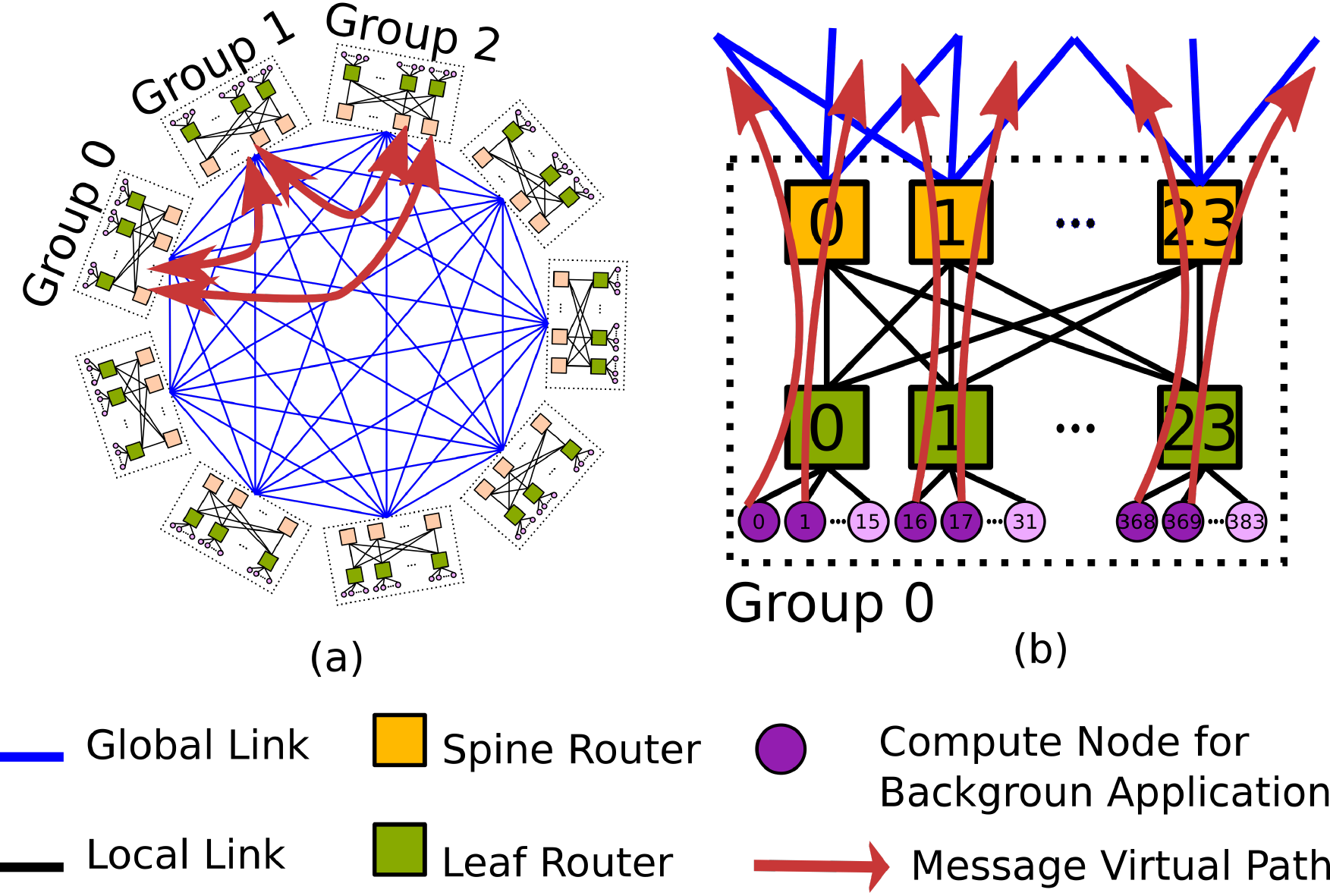}
	\caption{(a) Background traffic generated among three groups. (b) Group 0 with highlighted nodes holding background application processes}
	\label{fig:backnoise}
\end{figure}

Figure \ref{fig:backnoise}(a) gives an overview of the generated background traffic. Red arrows are virtual paths which background traffic follows. 
Our Dragonfly+ system has 48 global links between any two groups, thus for most of the experiments, we place the same number of background processes in each group with one process per node. As shown in Figure \ref{fig:backnoise}(b), the 48 background application nodes are equally distributed over 24 leaf routers. 
Background application generates inter-group messages between the source and destination nodes of the same local node ID but located in different groups. For example, node 0 in Group 0 only sends and receives messages from node 0 in Group 1 and Group 2. 


\begin{table}[htbp]
\begin{tabular}{crrr}
\hline
\multicolumn{1}{c|}{\multirow{2}{*}{Msg. Interval}} & \multicolumn{3}{c}{Comm. Intensity} \\ \cline{2-4} 
\multicolumn{1}{c|}{} & \multicolumn{1}{c|}{\begin{tabular}[c]{@{}c@{}}Underutilized\\ TGLL=50\%\end{tabular}} & \multicolumn{1}{c|}{\begin{tabular}[c]{@{}c@{}}Near-saturated\\ TGLL=90\%\end{tabular}} & \multicolumn{1}{c}{\begin{tabular}[c]{@{}c@{}}Overloaded\\ TGLL=130\%\end{tabular}} \\ \hline
\multicolumn{1}{c|}{1\textmu s} & \multicolumn{1}{r|}{2,340} & \multicolumn{1}{r|}{4,212} & 6,084 \\
\multicolumn{1}{c|}{10\textmu s} & \multicolumn{1}{r|}{23,400} & \multicolumn{1}{r|}{42,120} & 60,840 \\
\multicolumn{1}{c|}{100\textmu s} & \multicolumn{1}{r|}{234,000} & \multicolumn{1}{r|}{421,200} & 608,400 \\
\hline
\multicolumn{4}{c}{}
\end{tabular}
\caption{Background application configuration. The cells present message size in bytes under different message interval and communication intensity configurations.}
\label{tab:noiseload}
\end{table}


In this study we mainly focus on background application with three communication intensities: {\it under-utilized background} (TGLL=\\50\%), {\it near-saturated background (TGLL=90\%)} and {\it overloaded background (TGLL=130\%)}. These numbers are obtained by various combination of background application message sizes and message intervals.
Table \ref{tab:noiseload} presents background message sizes under different combinations of message intervals and communication intensities.

\subsection{Target Application} \label{sec:target_applicaton}

Our application interference analysis focuses on the performance of target application. In this study, the target application is a 2,304-process (6 group) MPI job with one process per node with four communication patterns and two job placement policies.

\subsubsection{Communication Pattern} {\text{ }} 

\textbf{Uniform Random (UR):}
In this pattern, each MPI process randomly chooses a destination process at each communication iteration. 
UR pattern has no global links preference as its communication destinations are randomly selected and tends to evenly distribute messages across the network. 
When the job size is as large as multiple of group size, most of the randomly selected source and destination nodes are located in different groups, which makes most of the network traffic into inter-group messages.

\textbf{3D Stencil:} 
In this pattern, MPI processes are organized as a 3D Cartesian grid and each process communicates with its six neighbors, two in each dimension. 3D Stencil is a common one-to-many communication pattern in HPC scientific workloads.

\textbf{Tornado:} 
In this pattern, each MPI process calculates its communication partner by adding a fixed offset value to its process ID. This offset is set to be equal to one group size in this study.

\textbf{Broadcasting:}
This pattern assigns a root process who broadcasts one message to all other processes at each communication iteration. Broadcasting is commonly used in MPI jobs for data synchronization.
It is chosen as a representative of one-to-all communication type.
In order to make the analysis of broadcasting pattern thorough and deep, we extend the background application's TGLL to 260\%, 390\%, and 520\% using the equation (\ref{eq:tgll}) and adding additional 3 background application nodes per router in the background application groups to achieve such TGLLs.

\newpage
\subsubsection{Job Placement Method} {\text{ }} 

\textbf{Contiguous placement:}
The 2,304 MPI processes of the target application are allocated in a contiguous manner in Group 3 to 8. In other words, the target application is isolated from the background application as they do not share any group.

\textbf{Random placement:}
The 2,304 MPI processes of the target application are allocated onto the system in a random manner from Group 0 to 8. We use the same random placement layout for all experiments under this placement. Randomize result shows that 29.6\% target application processes are placed in groups 0, 1 and 2. Hence the target application shares some groups with the background application. 

For the broadcasting pattern, we differentiate random placement into two cases:
\begin{itemize}
    \item Random placement with broadcasting root \textbf{\textit{outside}} background application groups 
    \item Random placement with broadcasting root \textbf{\textit{in}} background application groups. 
\end{itemize}

\begin{table}[htbp]
\begin{tabular}{c|ccc}
\hline
\multirow{2}{*}{Msg. Size} & \multicolumn{3}{c}{Comm. Intensity} \\ \cline{2-4} 
 & \multicolumn{1}{c|}{\begin{tabular}[c]{@{}c@{}}Underutilized\\ (TGLL\\ = 21\% - 28\%)\end{tabular}} & \multicolumn{1}{c|}{\begin{tabular}[c]{@{}c@{}}Near-saturated\\ (TGLL\\ = 73\% - 85\%)\end{tabular}} & \begin{tabular}[c]{@{}c@{}}Overloaded\\ (TGLL\\ \textgreater 100\%)\end{tabular} \\ \hline
4KB & \multicolumn{1}{c|}{3\textmu s} & \multicolumn{1}{c|}{1\textmu s} & 0.5\textmu s \\
512KB & \multicolumn{1}{c|}{450\textmu s} & \multicolumn{1}{c|}{150\textmu s} & 100\textmu s \\
4MB & \multicolumn{1}{c|}{4ms} & \multicolumn{1}{c|}{1ms} & 0.7ms \\ \hline
\multicolumn{4}{c}{} \\
\end{tabular}
\caption{Target application configuration. The cells present message intervals for UR/Stencil/Tornado under different message size and communication intensity configurations.}
\label{tab:gcload}
\end{table}


In this study, we analyze three target application message sizes (4KB, 512KB, and 4MB) under different communication intensities: {\it underutilized, near-saturated, and overloaded}. Table \ref{tab:gcload} illustrates the target application configuration with cells indicating the message interval under different message size and communication intensity combinations for UR, 3D stencil and tornado patterns.
In order to represent more realistic applications, broadcasting intervals are chosen long enough such that a new broadcasting iteration is initiated after all the previous broadcasting messages reach their destination.

Our interference evaluation focuses on both {\it intra-job interference} and {\it inter-job interference}. A {\it baseline} case is set where the target application is executed without other jobs in the system (i.e., an ideal non-interference scenario). {\it Message latencies} of the target application running under various background application intensities are used to study intra- and inter-job interference. 

For the UR, 3D stencil and tornado pattern, we investigate 2 different job placements, 3 message sizes, and 3 message intervals, under 4 background application loads: baseline, under-utilized (50\% TGLL), near-saturated (90\% TGLL), and overloaded (130\% TGLL). 
For the broadcasting pattern, experiments are designed with heavier, ranging from baseline to 520\% TGLL background application loads under 3 job placement scenarios, namely contiguous placement, random placement where the broadcasting root is located outside the background application groups (group 0-2), and random placement where the broadcasting root located in the background application groups.

\begin{figure*}[htbp]
\centering
\begin{tabular}{ccc}
\multicolumn{3}{c}{\includegraphics[width=0.7\textwidth]{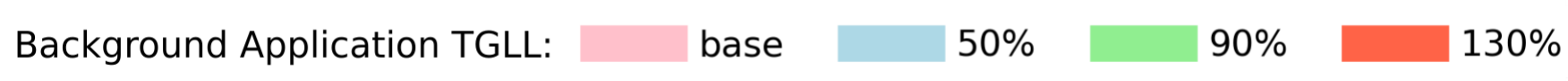}} \\
\includegraphics[width=0.3\textwidth]{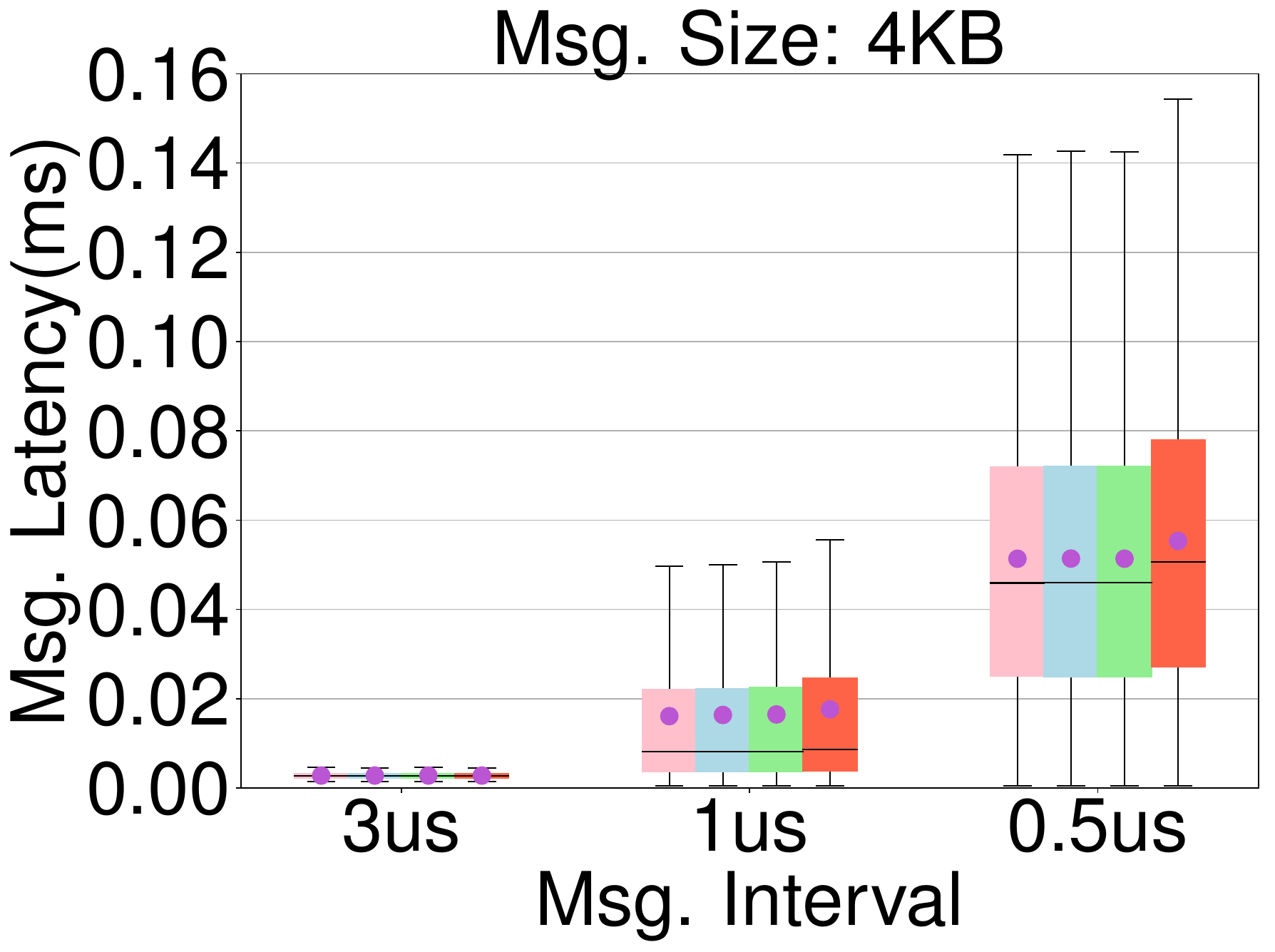} &
\includegraphics[width=0.3\textwidth]{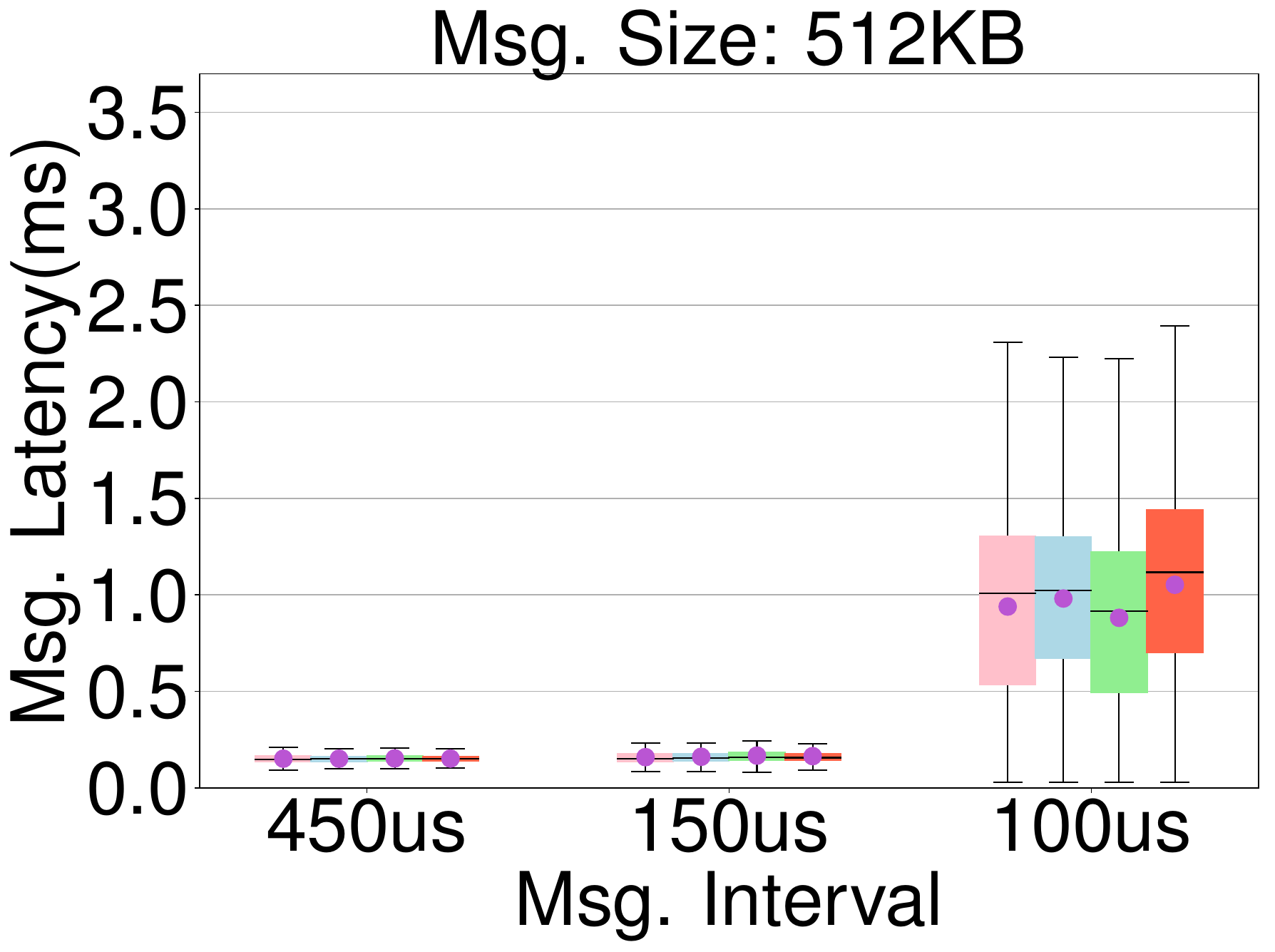} &
\includegraphics[width=0.3\textwidth]{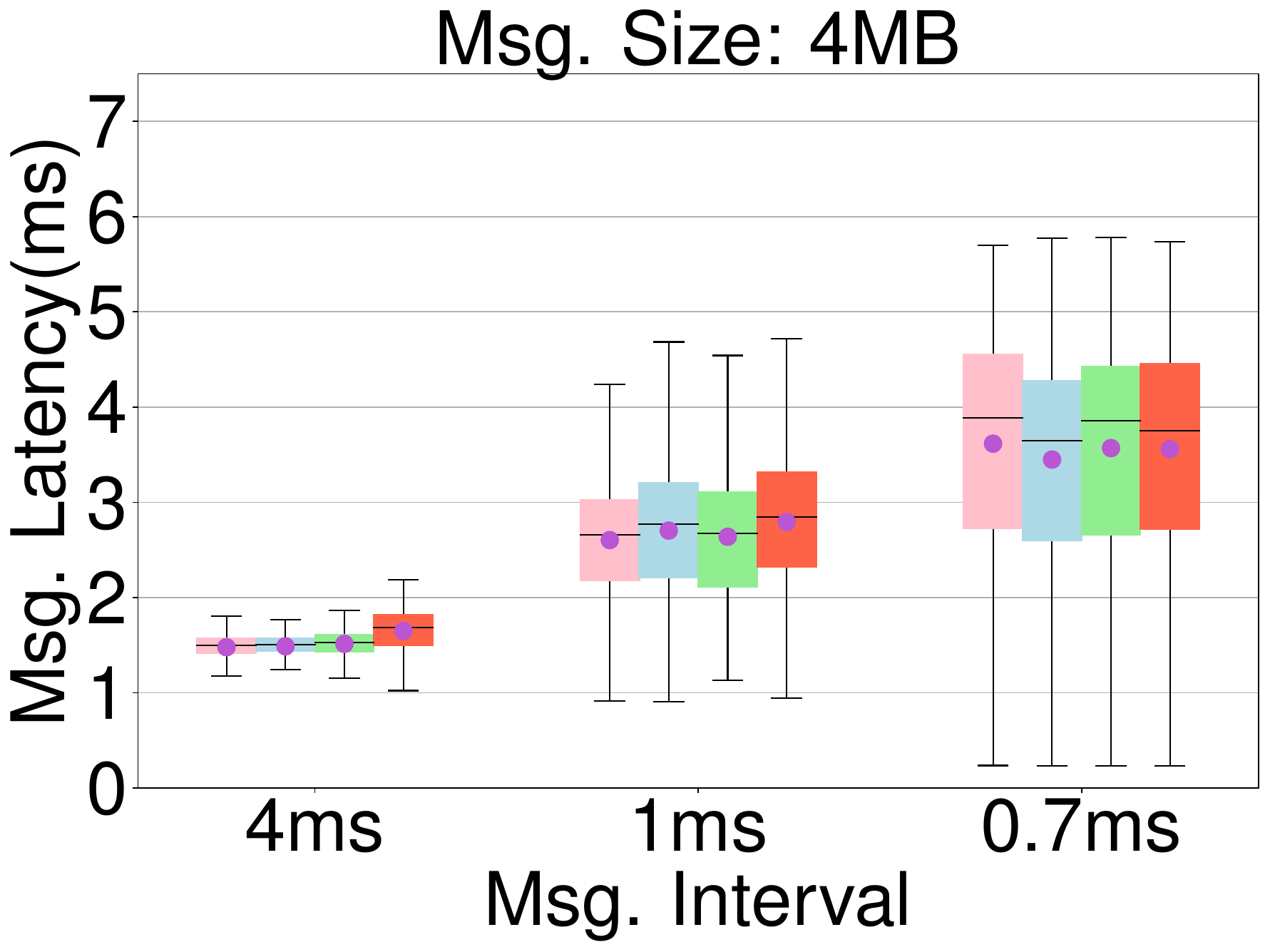} \\
(a) & (b) &(c) \\
\multicolumn{3}{c}{Contiguous placement without group overlapping} \\
\\
\includegraphics[width=0.3\textwidth]{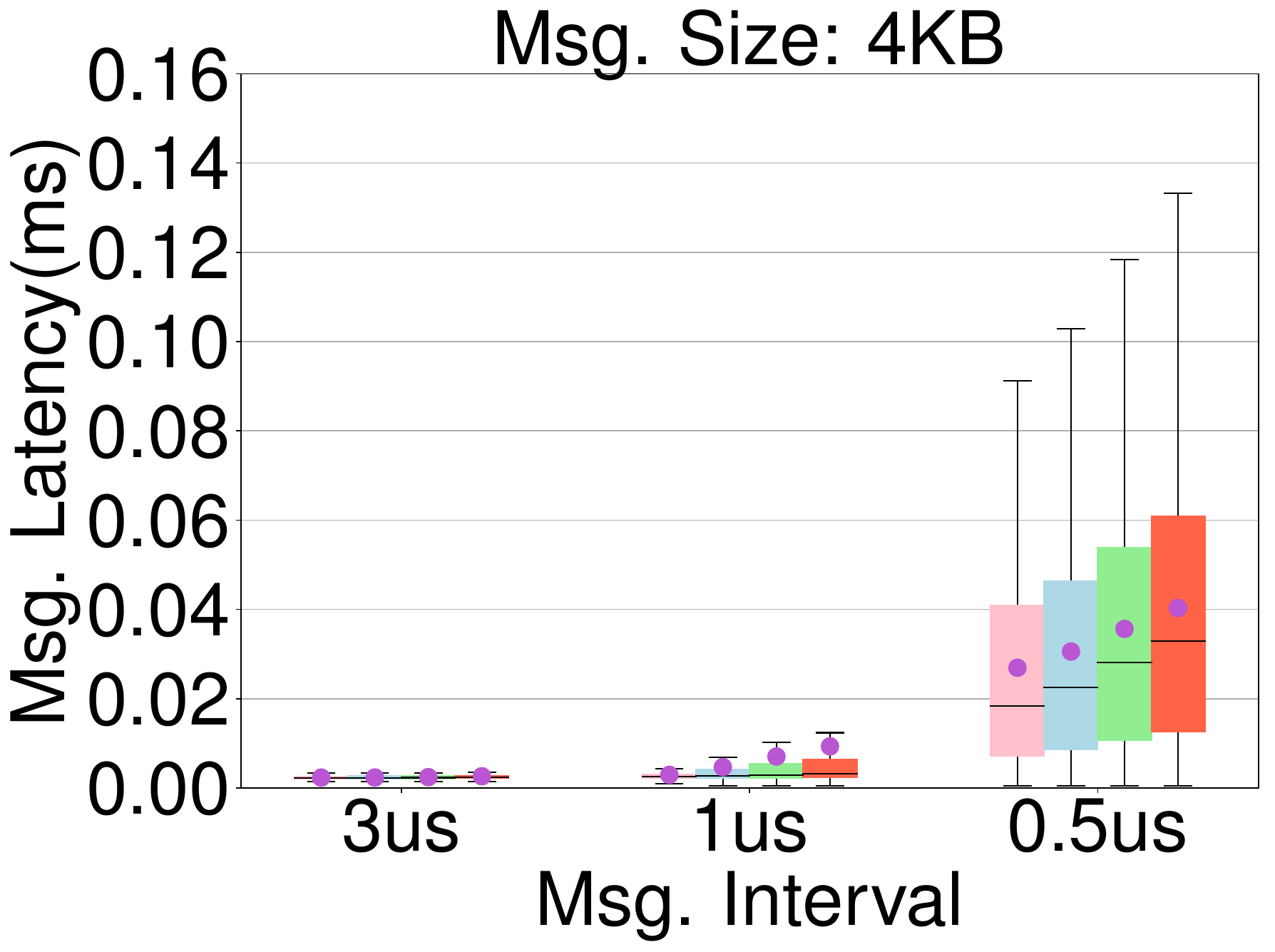} &
\includegraphics[width=0.3\textwidth]{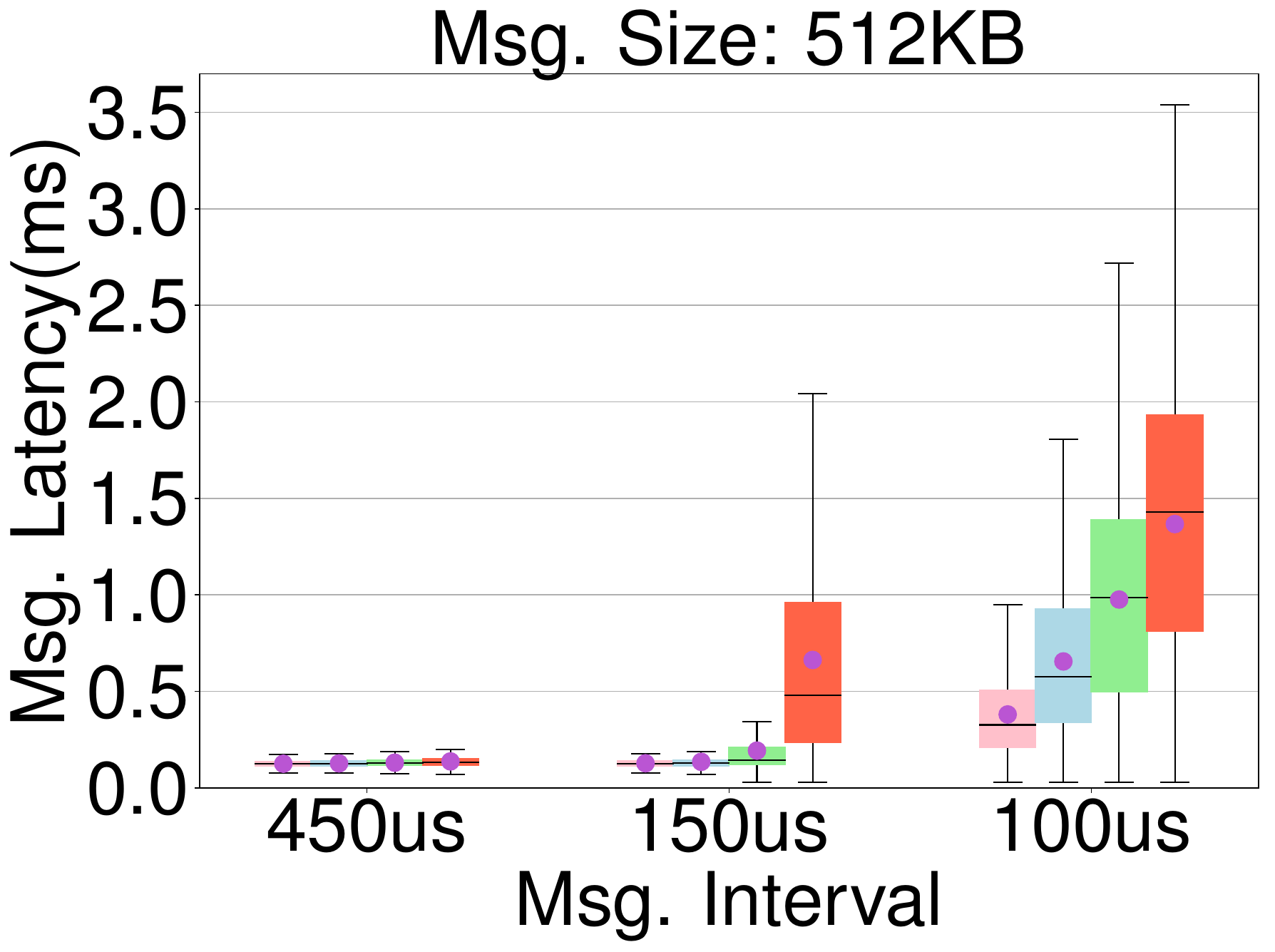} &
\includegraphics[width=0.3\textwidth]{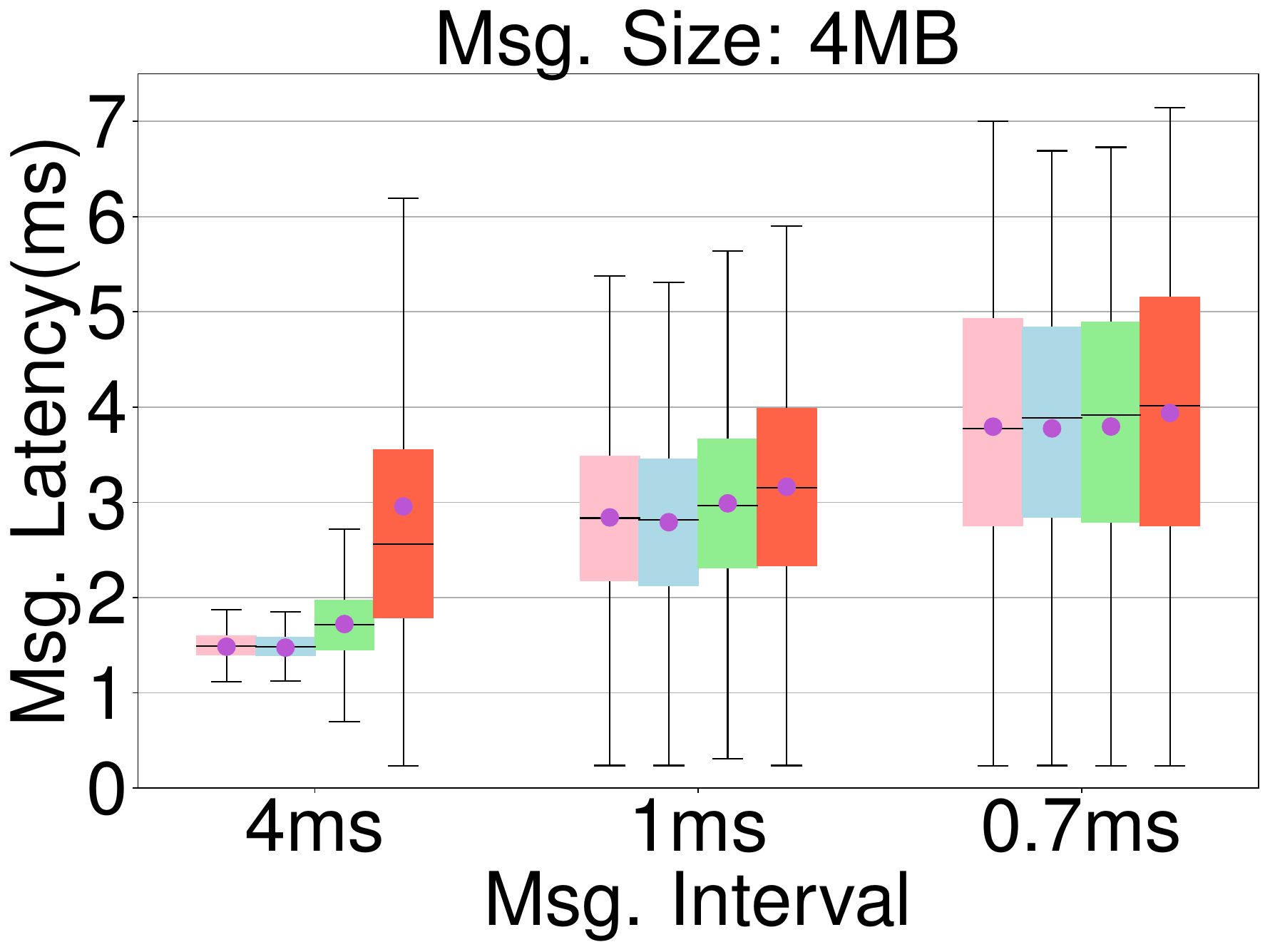} \\
(d) & (e) & (f) \\
\multicolumn{3}{c}{Random placement} \\
\end{tabular}
\caption{Message latency of the target application with Uniform Random (UR) communication pattern. 
The top row depicts the use of contiguous job placement of the target application without group overlapping with the background application. The bottom row shows the result of the target application shares groups with the background application under the random placement. Message latency under different background application loads is identified by its color, where baseline is the target application executed solely on the system.}
\label{fig:ur}
\end{figure*}


\begin{figure*}[htbp]
\centering
\begin{tabular}{ccc}
\multicolumn{3}{c}{\includegraphics[width=0.7\textwidth]{pic/legend.pdf}} \\
\includegraphics[width=0.3\textwidth]{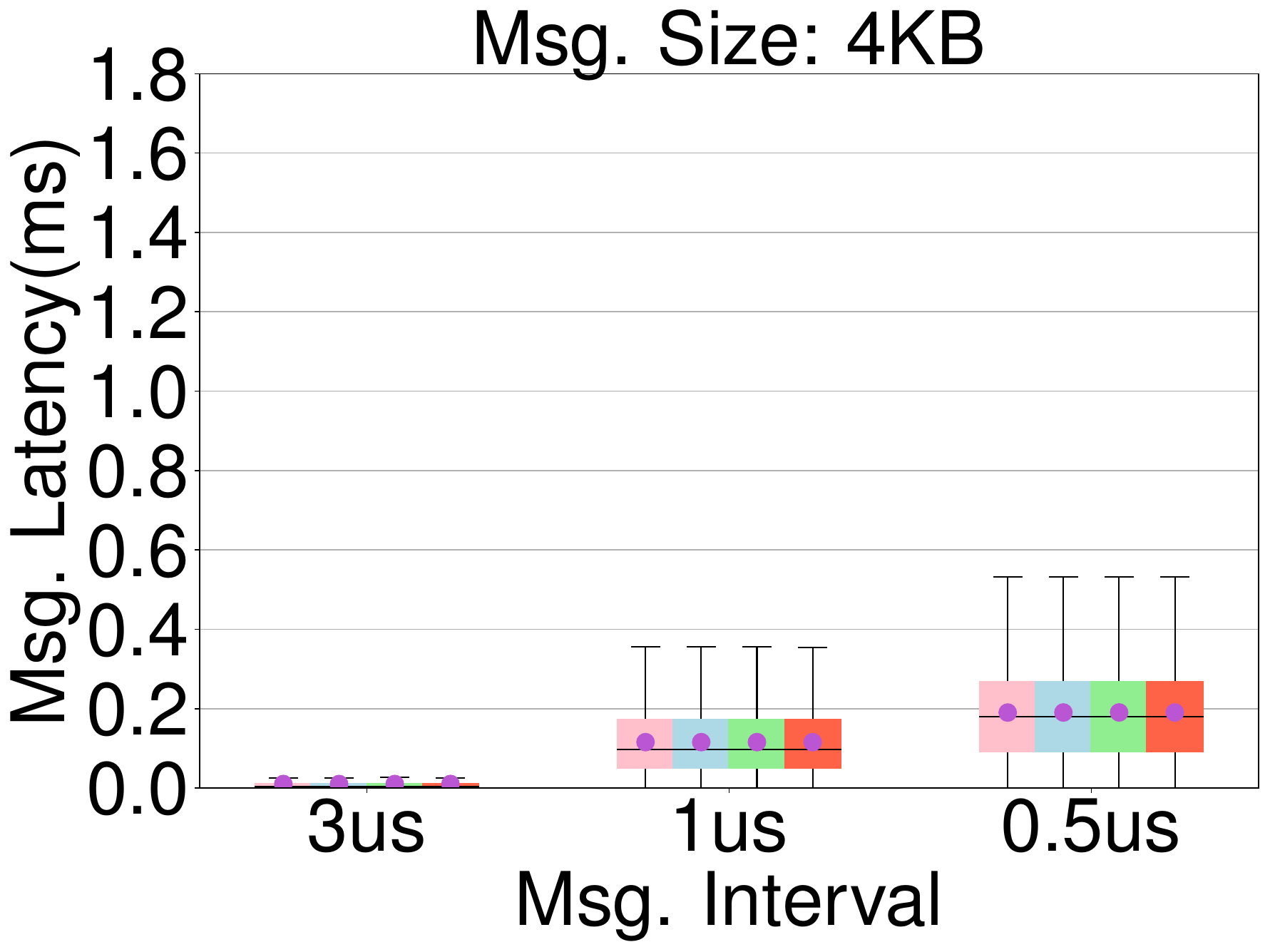} &
\includegraphics[width=0.3\textwidth]{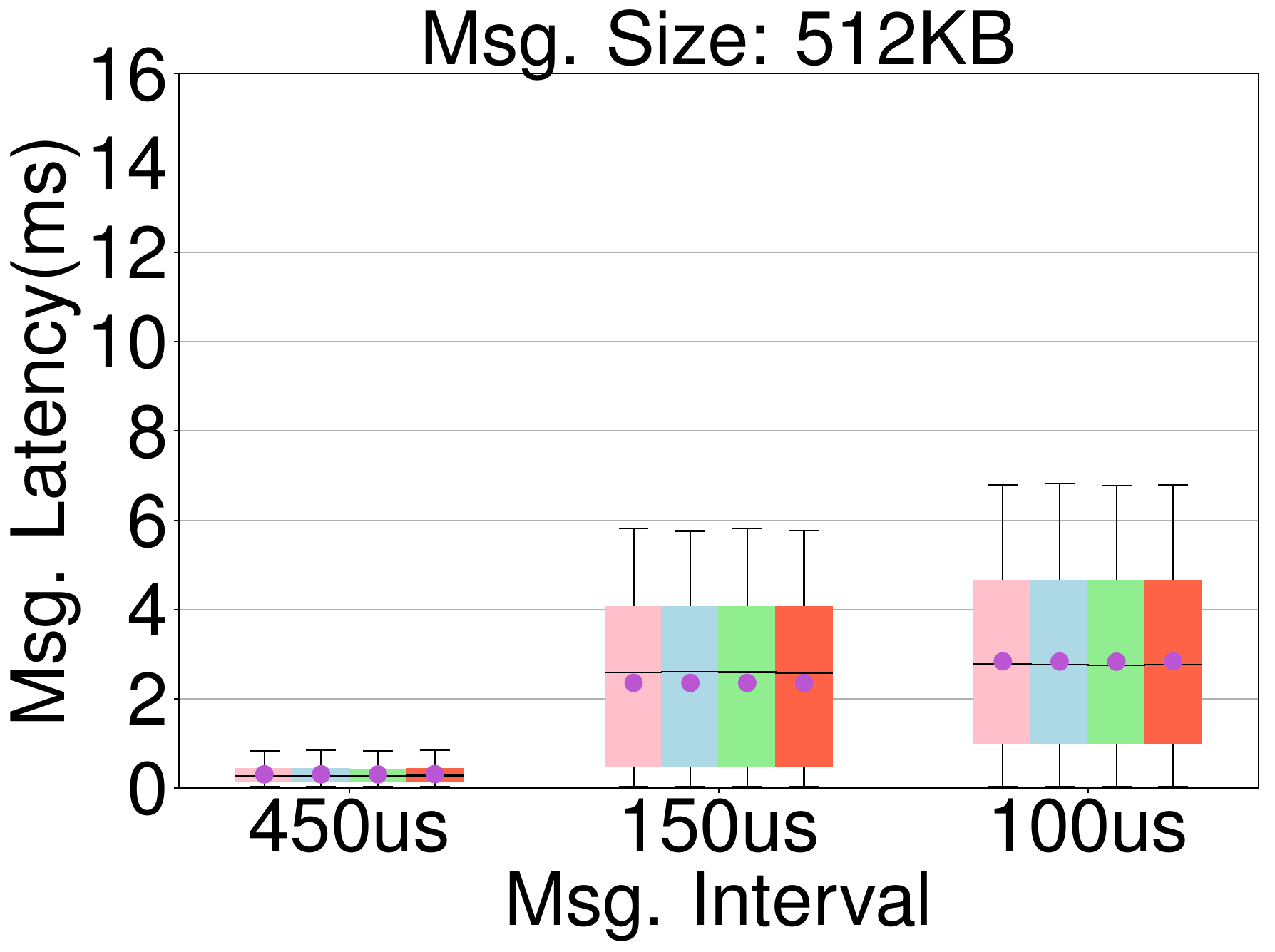} &
\includegraphics[width=0.3\textwidth]{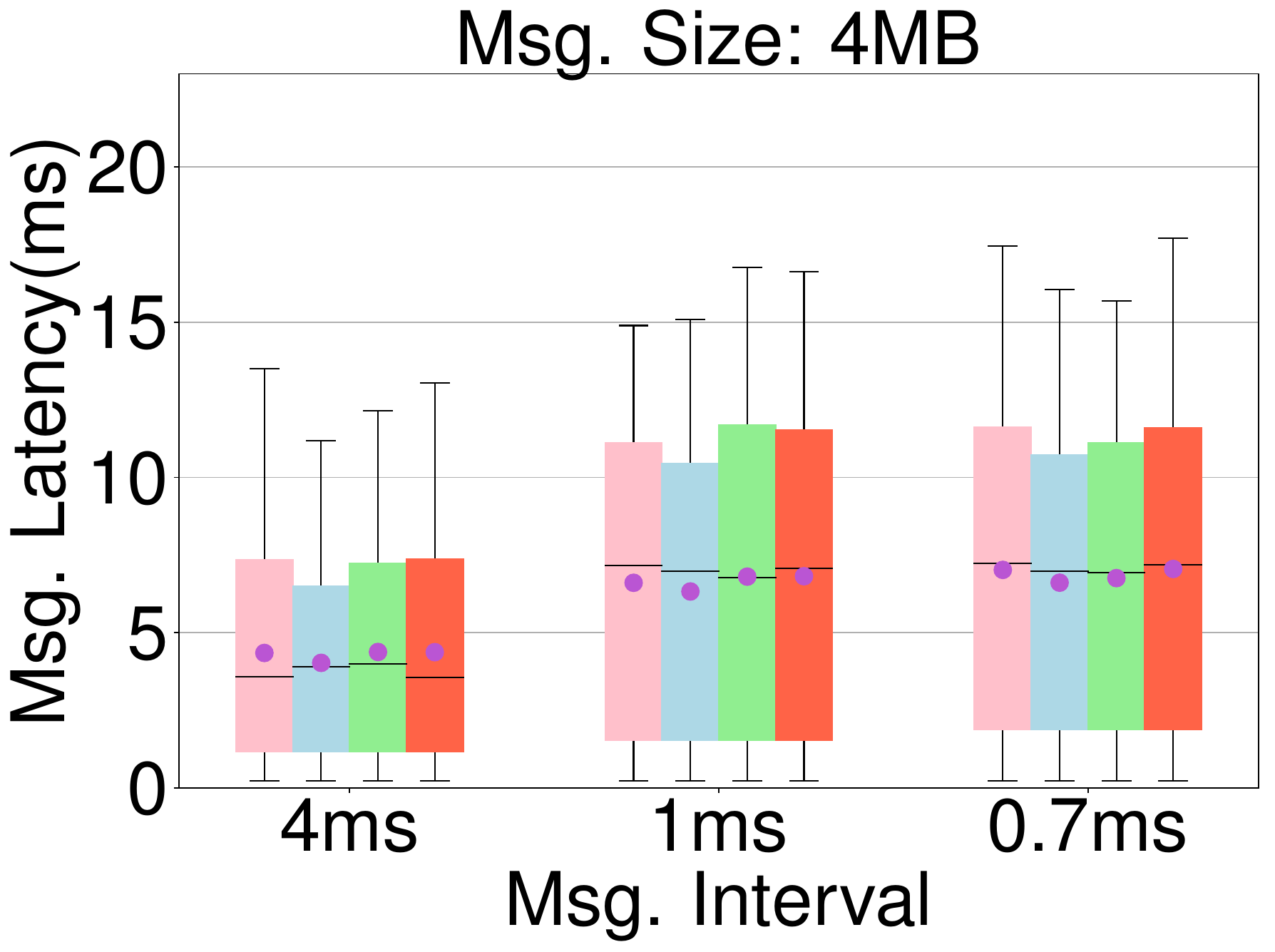} \\
(a) & (b) &(c) \\
\multicolumn{3}{c}{Contiguous placement without group overlapping} \\
\\
\includegraphics[width=0.3\textwidth]{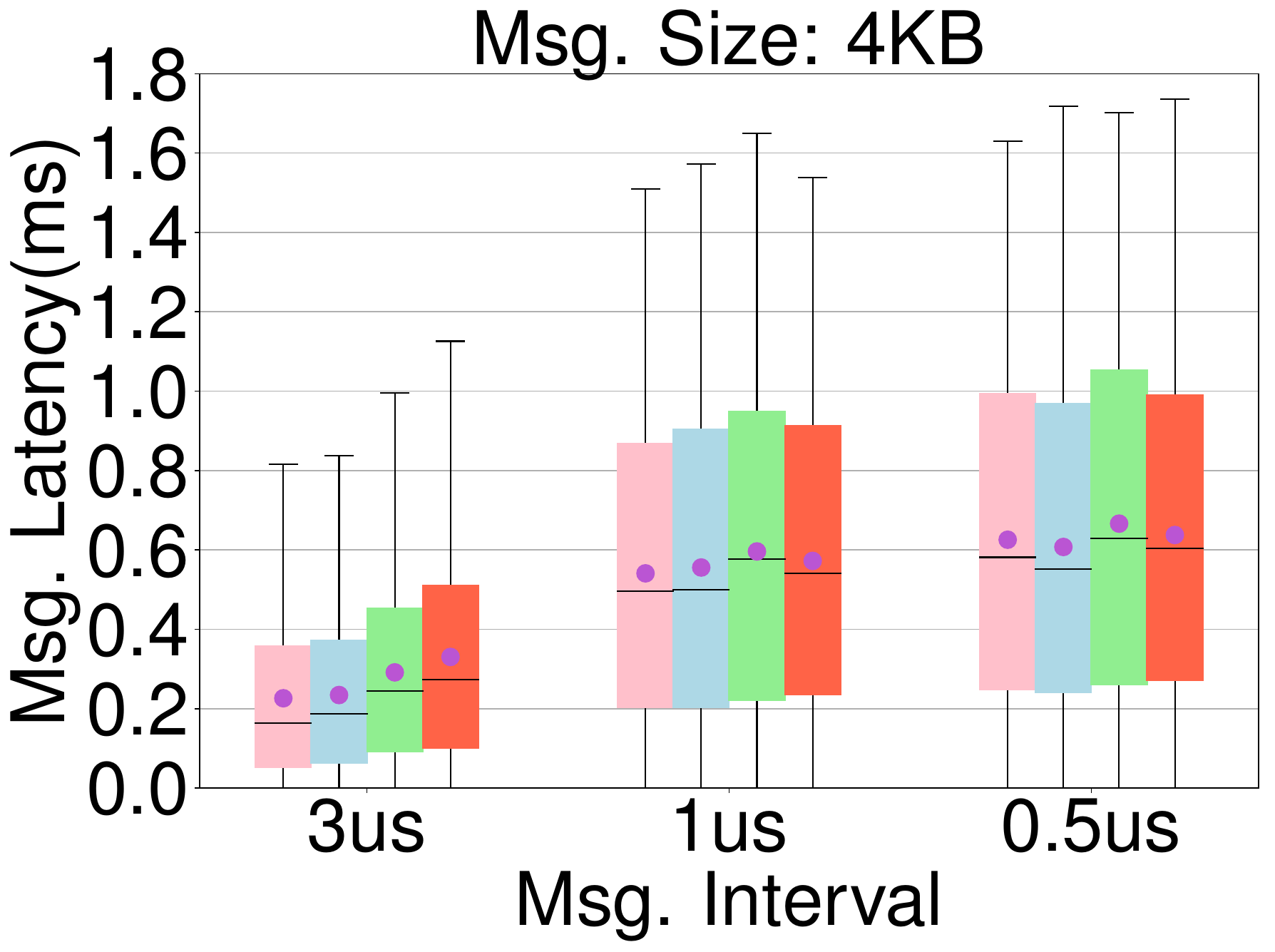} &
\includegraphics[width=0.3\textwidth]{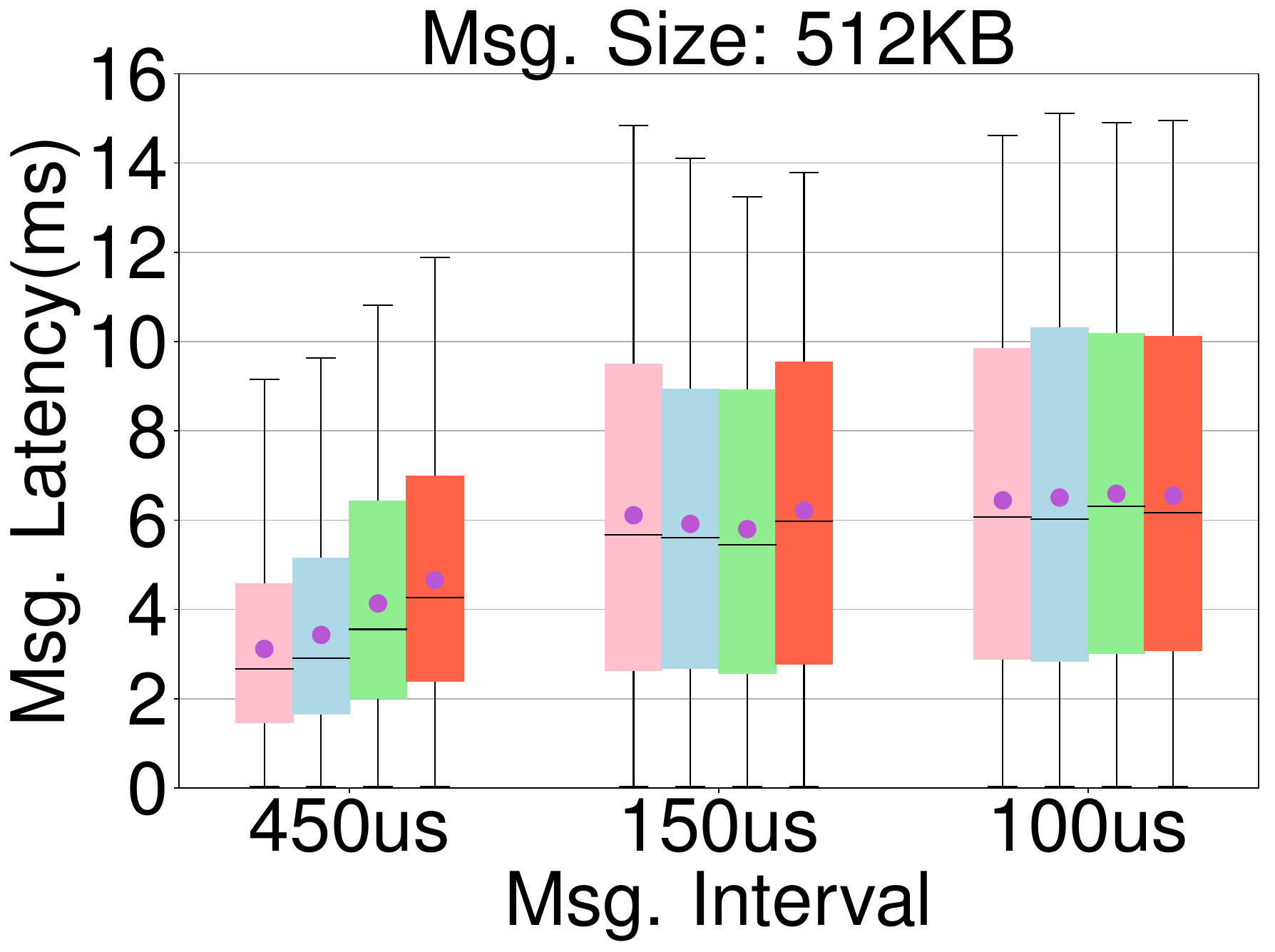} &
\includegraphics[width=0.3\textwidth]{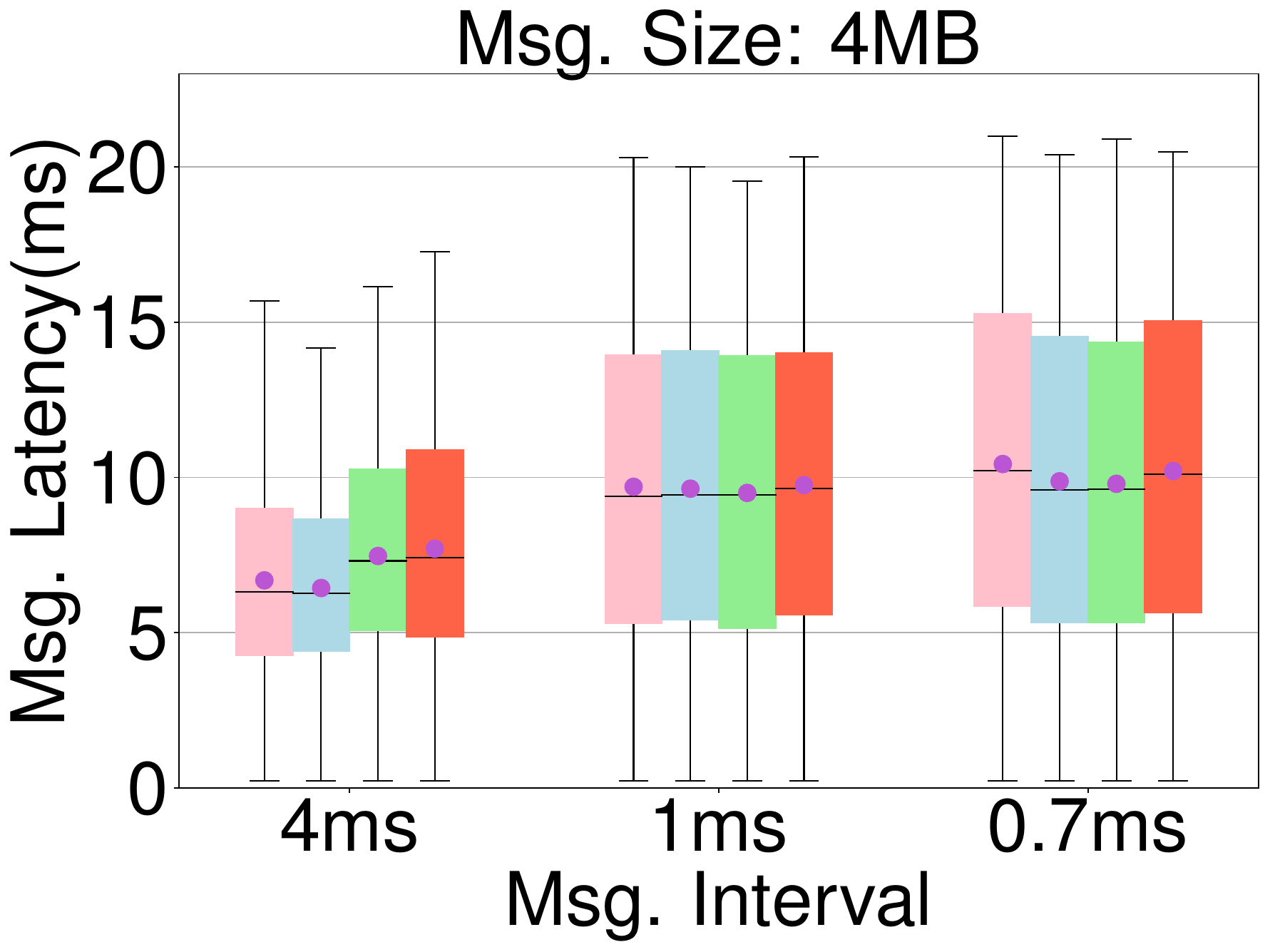} \\
(d) & (e) & (f) \\
\multicolumn{3}{c}{Random placement} \\
\end{tabular}
\caption{Message latency of the target application with 3D stencil communication pattern.}
\label{fig:stencil}
\end{figure*}


\begin{figure*}[htbp]
\centering
\begin{tabular}{ccc}
\multicolumn{3}{c}{\includegraphics[width=0.7\textwidth]{pic/legend.pdf}} \\
\includegraphics[width=0.3\textwidth]{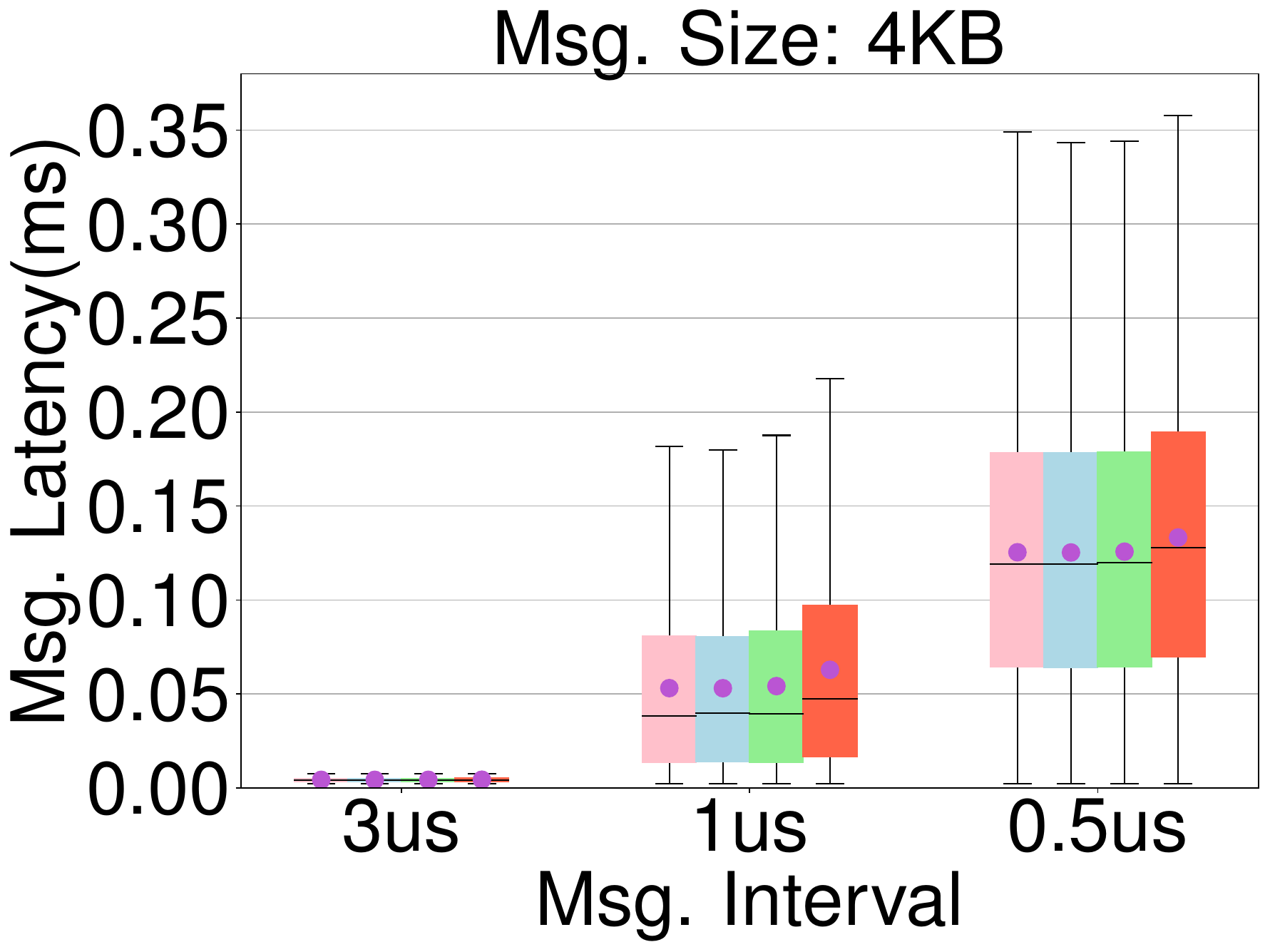} &
\includegraphics[width=0.3\textwidth]{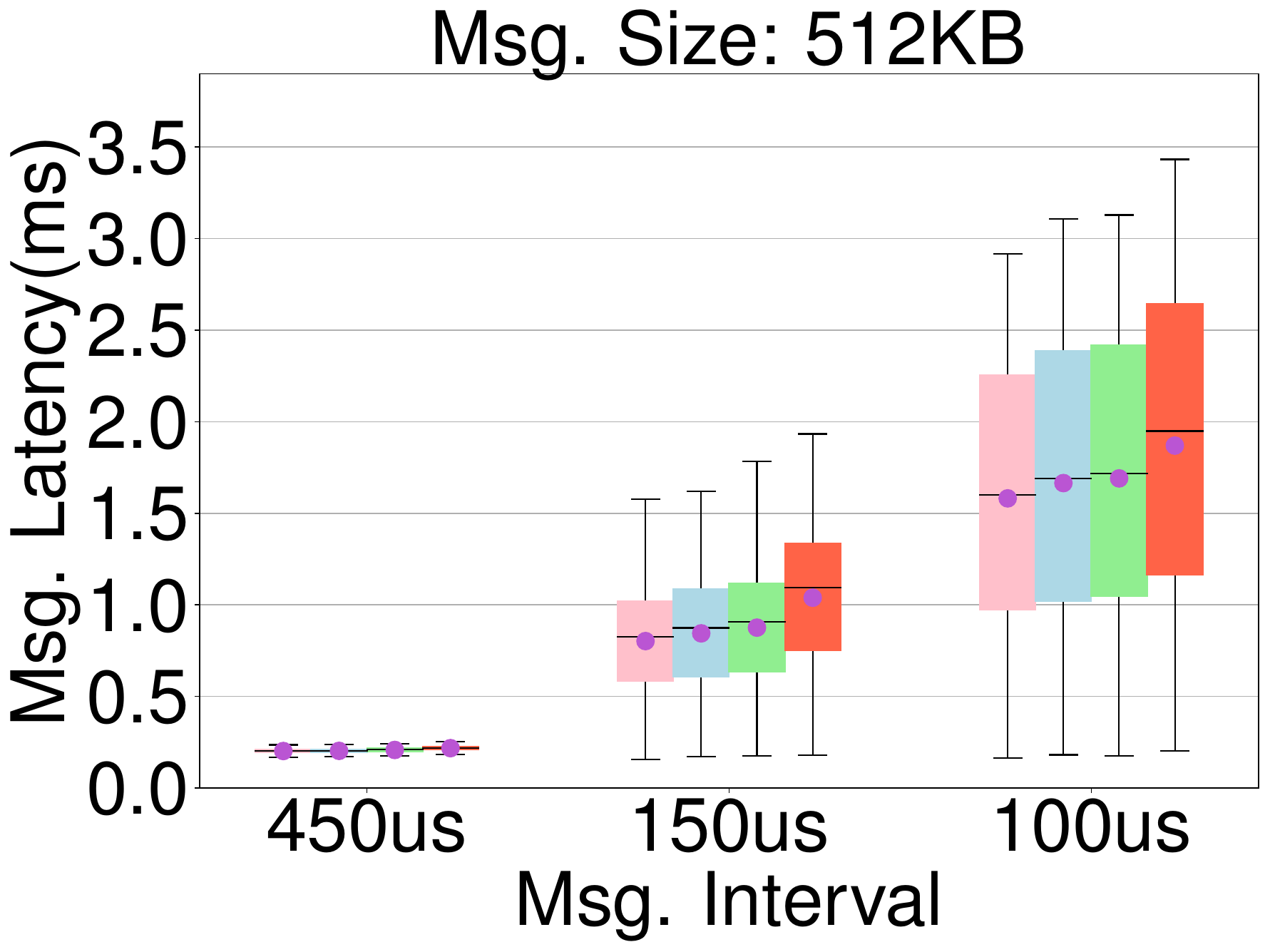} &
\includegraphics[width=0.3\textwidth]{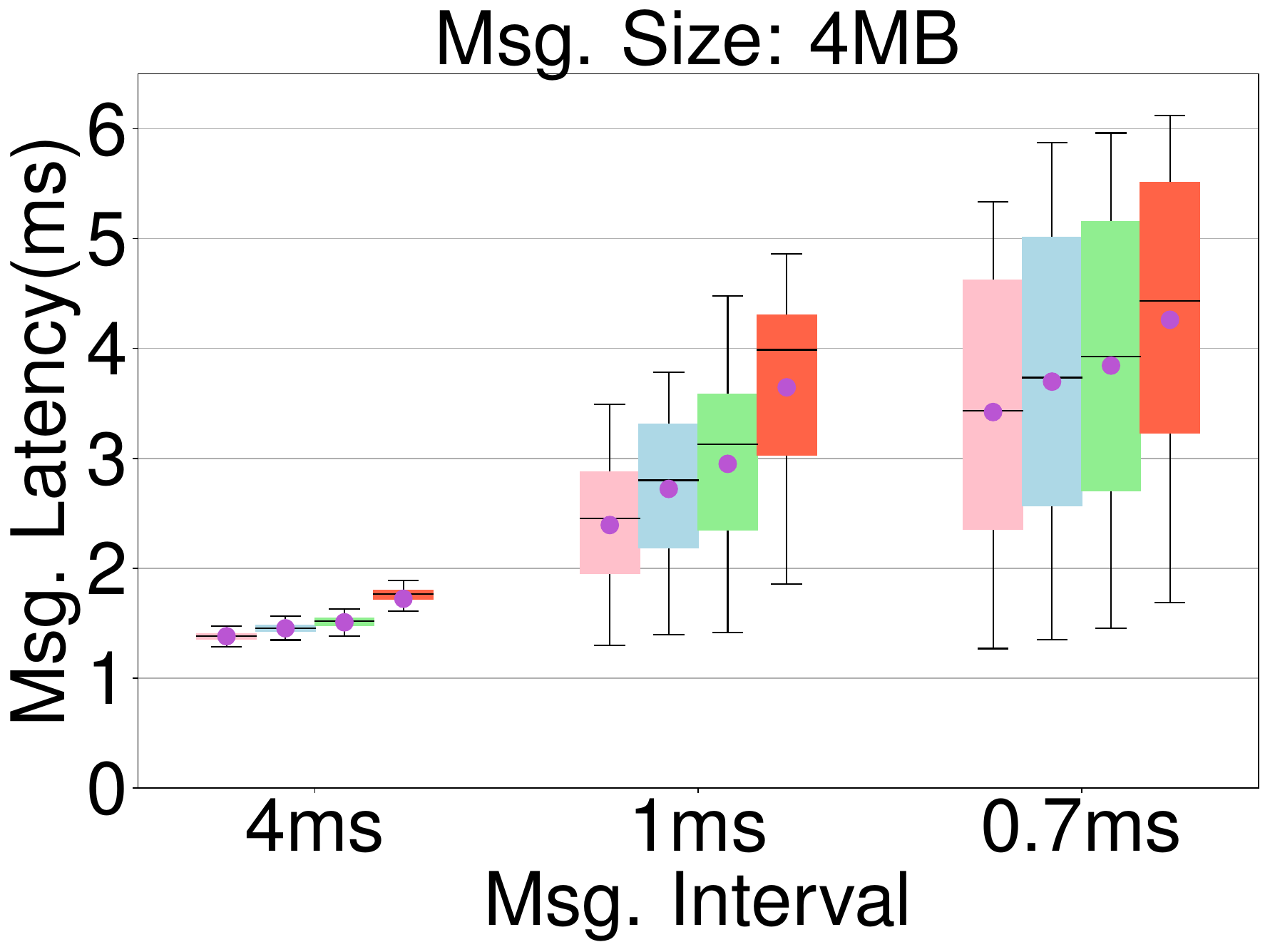} \\
(a) & (b) &(c) \\
\multicolumn{3}{c}{Contiguous placement without group overlapping} \\
\\
\includegraphics[width=0.3\textwidth]{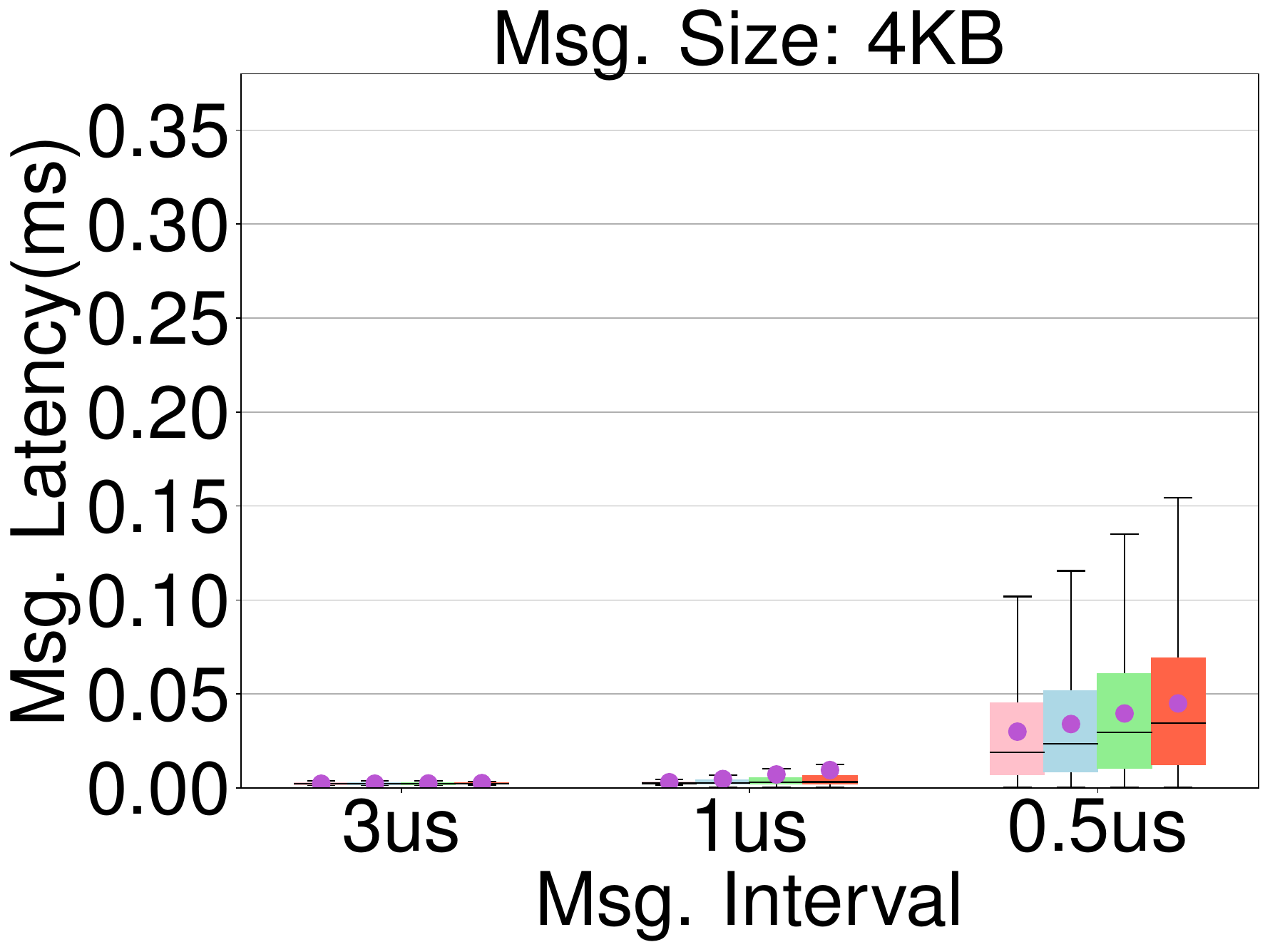} &
\includegraphics[width=0.3\textwidth]{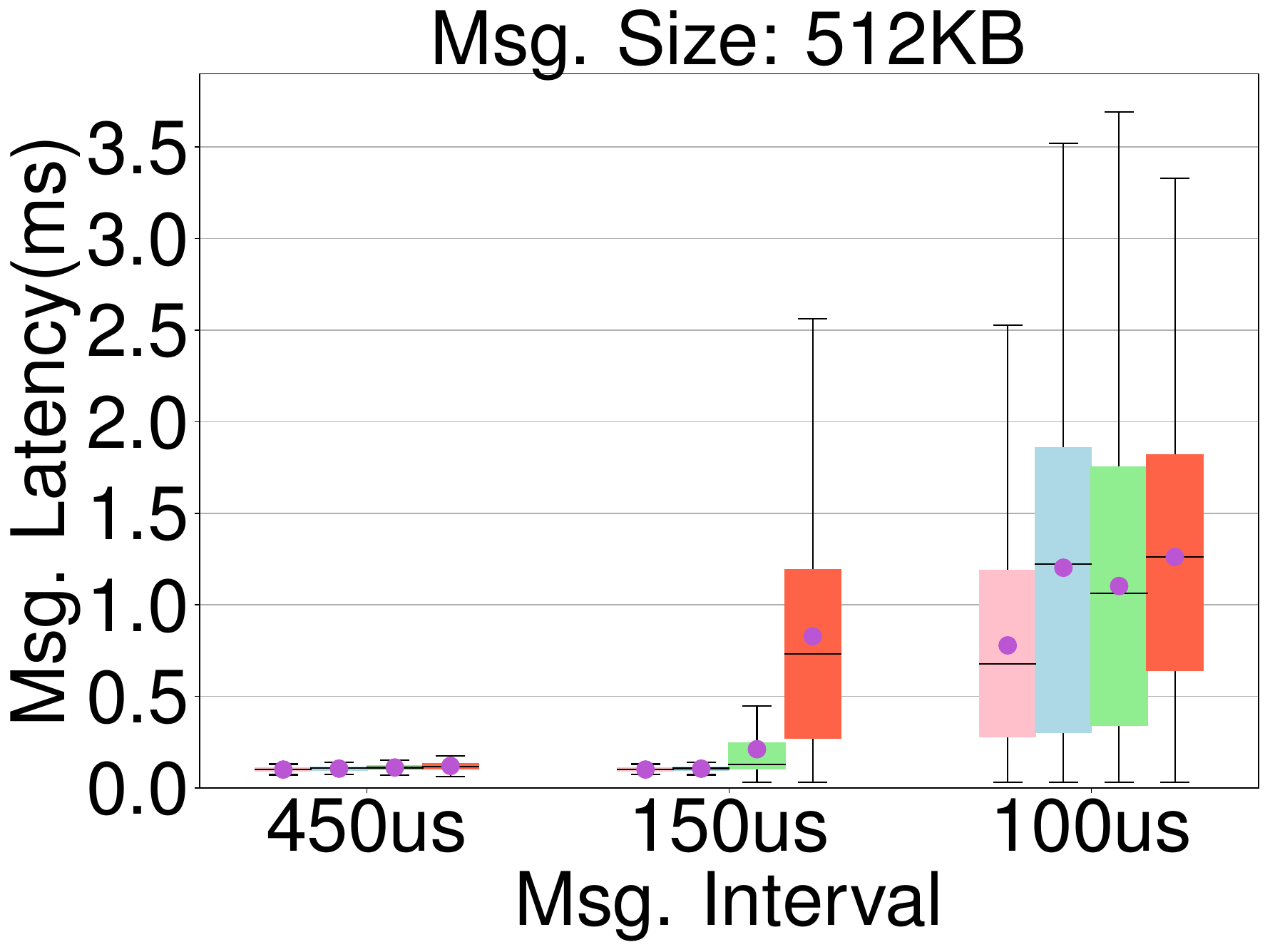} &
\includegraphics[width=0.3\textwidth]{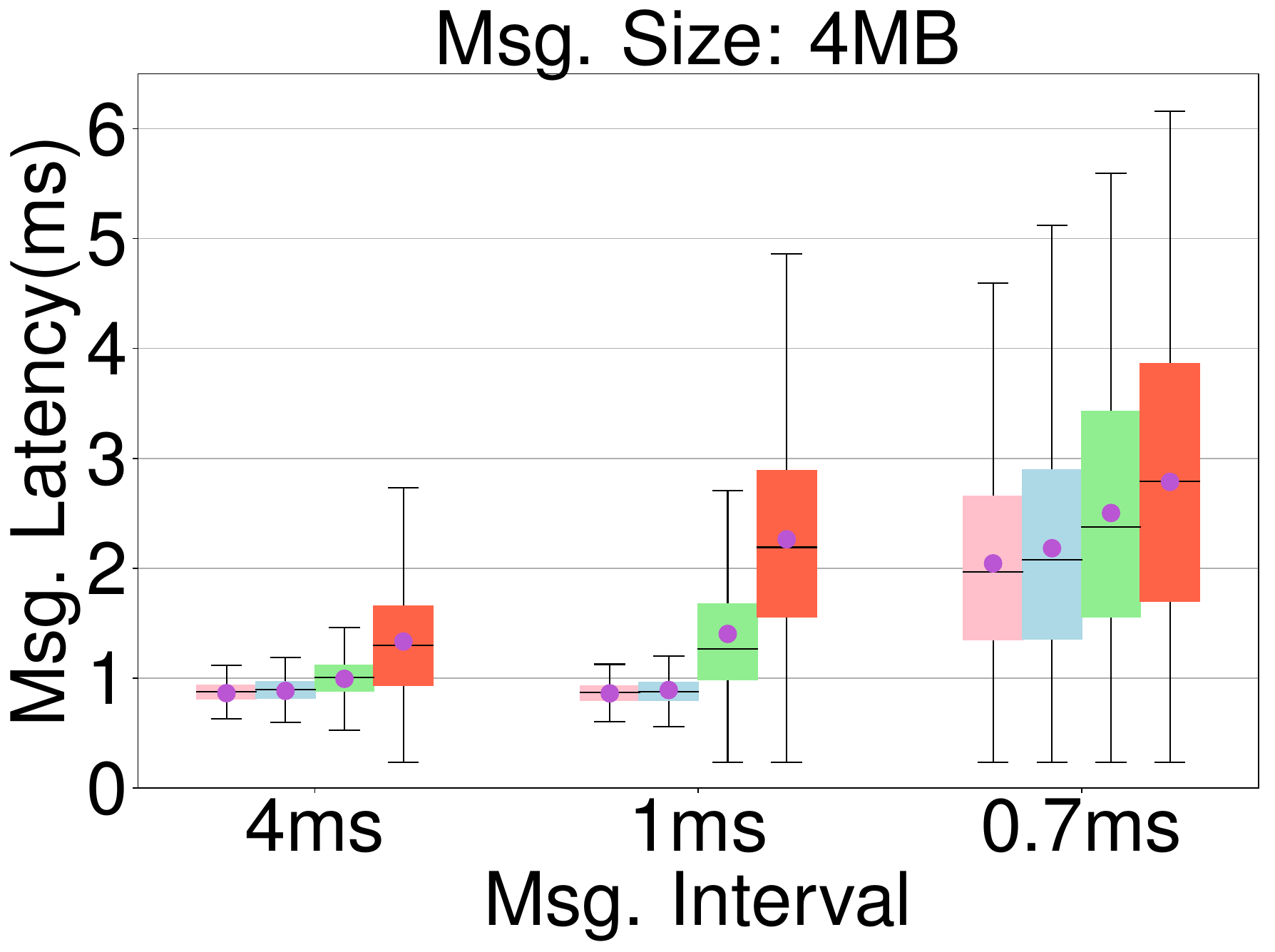} \\
(d) & (e) & (f) \\
\multicolumn{3}{c}{Random placement} \\
\end{tabular}
\caption{Message latency of the target application with tornado communication pattern.}
\label{fig:tornado}
\end{figure*}

\begin{figure}[htbp]
\centering
\includegraphics[width=1\linewidth]{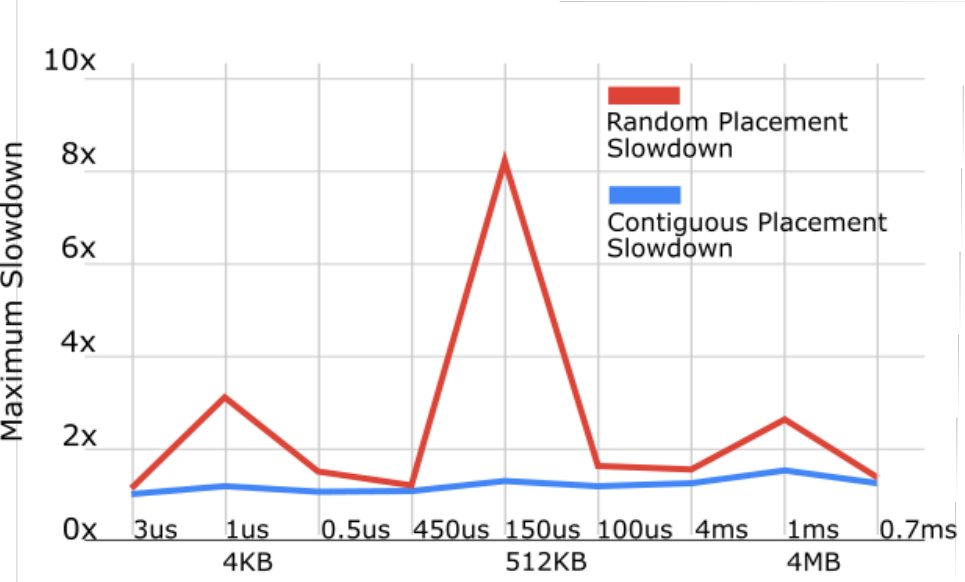}
\caption{Maximum slowdown of the average message latency of tornado pattern caused by inter-job interference}
\label{fig:tornado-interjob}
\end{figure}

\begin{figure*}[htbp]
\centering
\begin{tabular}{ccc}
\includegraphics[width=0.3\textwidth]{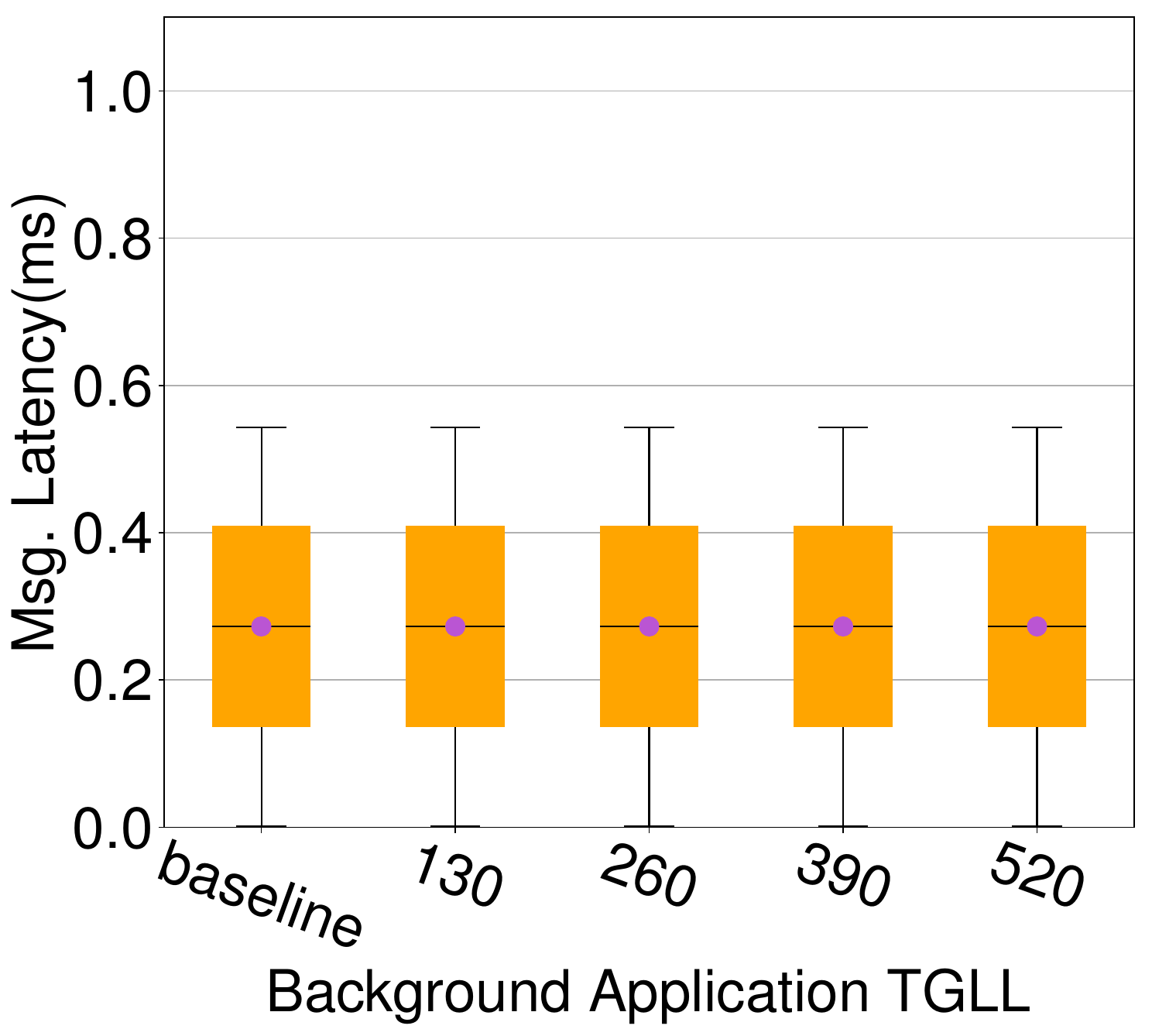} &
\includegraphics[width=0.3\textwidth]{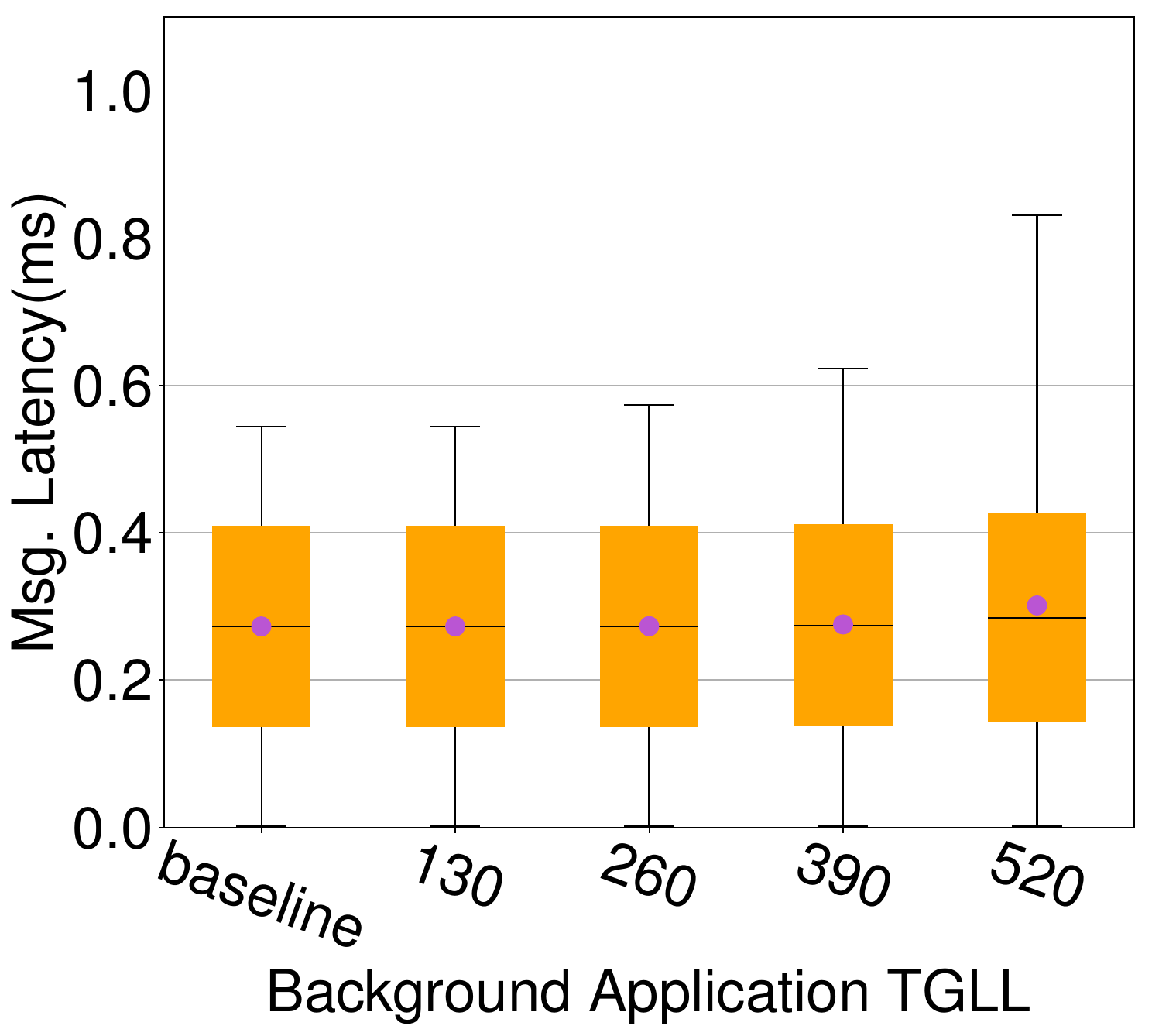} &
\includegraphics[width=0.3\textwidth]{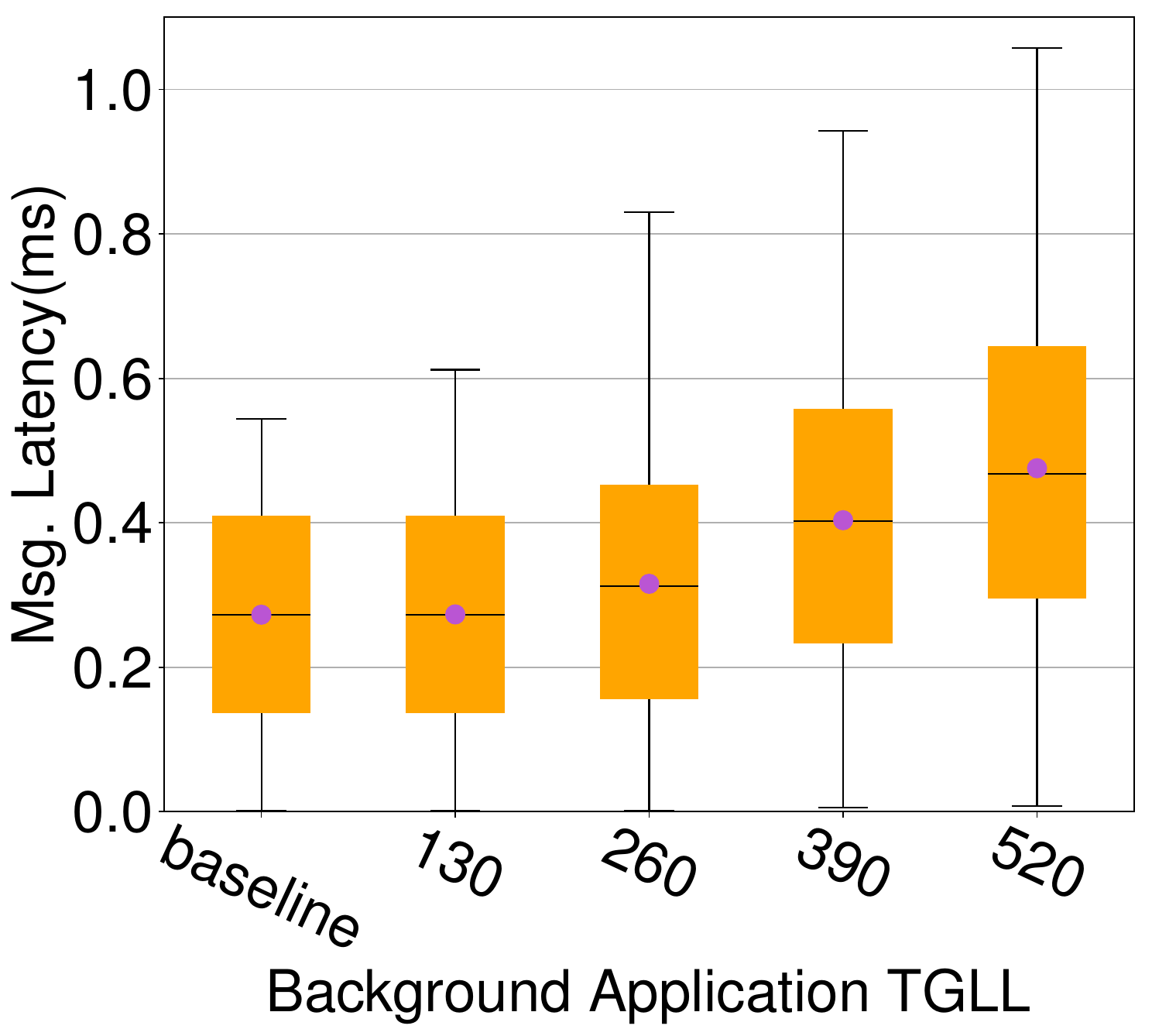} \\
\multicolumn{3}{c}{} \\
(a) Contiguous placement & (b) Random placement, broadcasting root & (c) Random placement, broadcasting root\\ 
& outside the background application groups & in the background application groups\\ 
\end{tabular}
\caption{4KB message latency of the target application with broadcasting communication pattern}
\label{fig:bc-inter}
\end{figure*}

\begin{figure*}[htbp]
\centering
\begin{tabular}{ccc}
\includegraphics[width=0.3\textwidth]{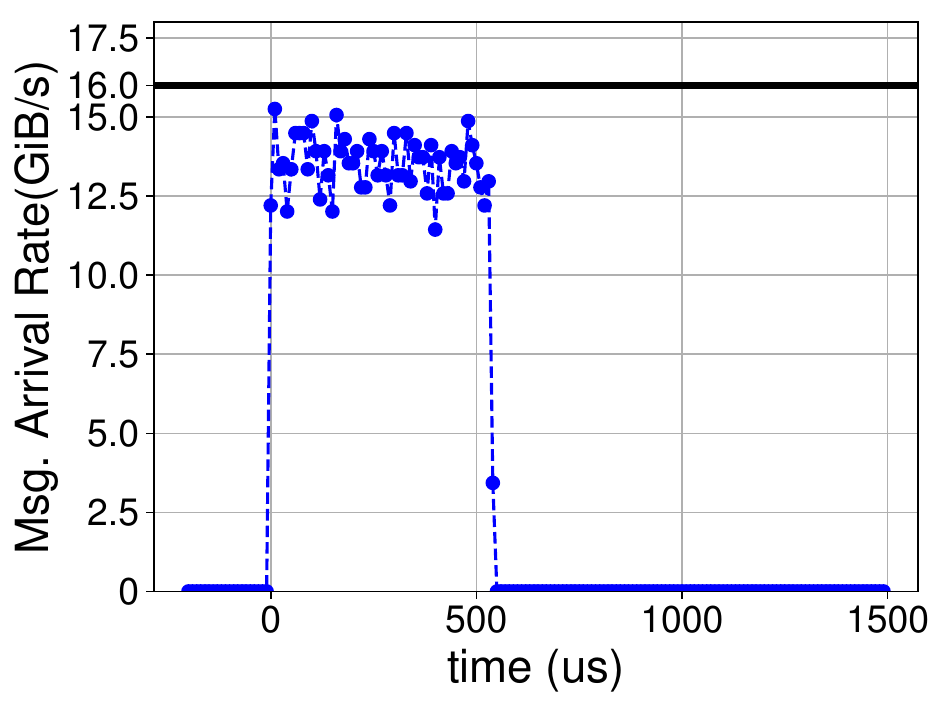} &
\includegraphics[width=0.3\textwidth]{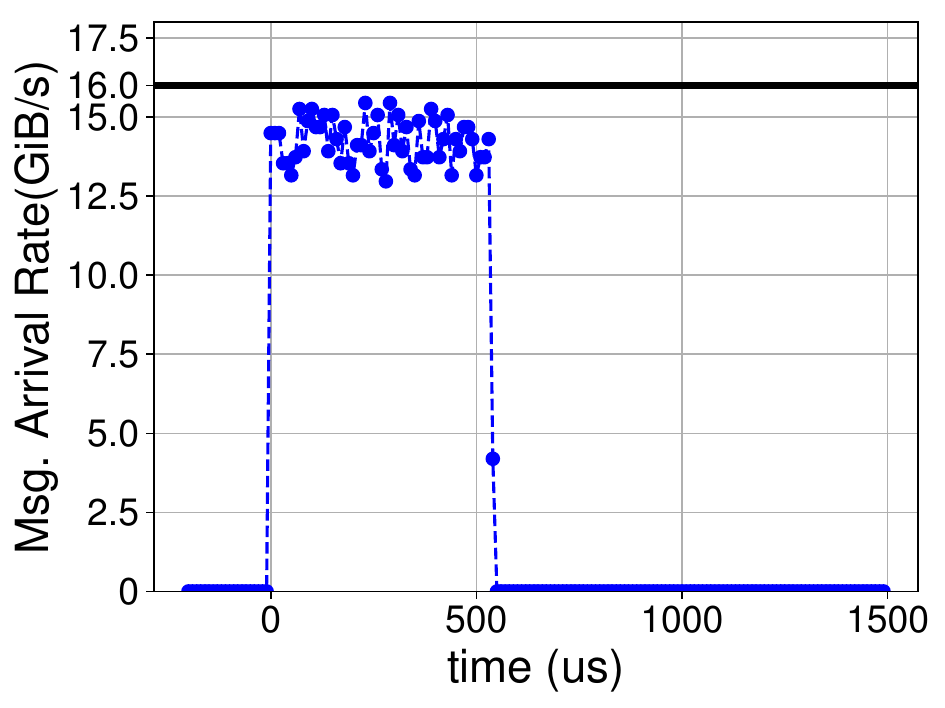} &
\includegraphics[width=0.3\textwidth]{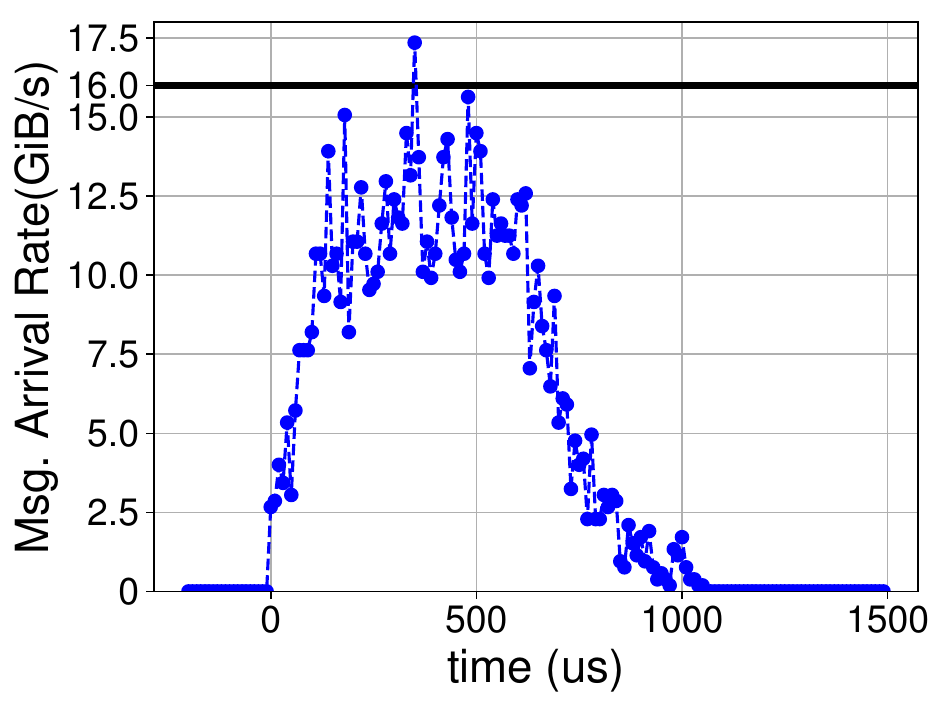} \\
(a) Contiguous placement & (b) Random placement, broadcasting root & (c) Random placement, broadcasting root\\ 
& outside the background application groups & in the background application groups\\ 
\end{tabular}
\caption{4KB message broadcasting inter-group message arrival rate at the broadcasting root group under the 520\% background application. The 16GiB/s node-to-router channel bottleneck is highlighted.}
\label{fig:bc-inject}
\end{figure*}

\section{Experimental Analysis} \label{sec:result}

In this section, we present application interference analysis on a 3,456-node Dragonfly+ system.
Application interference is a complicated problem influenced by various factors including background traffic and target application itself. In this study, we consider various background traffic intensities as listed in Table \ref{tab:noiseload} and target applications as listed in Table \ref{tab:gcload}. For each target application, we study two different job placement policies, namely contiguous and random placement, as described in Section \ref{sec:target_applicaton}.

\subsection{Uniform Random (UR)}

Figure \ref{fig:ur} presents {\it message latency} of the target application with UR communication pattern. The top three plots from (a) to (c) list the results with the contiguous placement where the target application does not share groups with the background application. The bottom plots from (d) to (f) present the results of the random placement where the target application shares groups with the background application.
In this experiment, we consider nine different configurations of message sizes and message intervals for the target application as listed in Table \ref{tab:gcload}. For each configuration of the target application, we simulate its execution under three background intensities (shown in Table \ref{tab:noiseload}), along with the ideal baseline with no external job interference. 
Each boxplot contains whiskers for the minimum, the 25\% percentile, the median, the 75\% percentile, and the maximum. The purple circle indicates the arithmetic average of the message latencies. 

\textbf{\textit{Baseline performance.}} 
The baseline results shown in the pink boxes indicate the ideal communication performance of the target application when it has an exclusive access of the system. 
Because there is no other jobs on the system, the baseline results help us understand the impact of job placement policies and the intra-job interference of the target application under different communication intensities.
The baseline performance shows that the random placement can help reduce the average message latency for small message size, but makes no important difference for large message size. 4KB message latency under contiguous placement is on average 2.8x longer than random placement, but 4MB message experience less than $1\%$ difference.

\textbf{\textit{Intra-job interference.}} 
The baseline message latencies of the target application in Figure \ref{fig:ur} show that shorter message interval (higher communication intensity) increases intra-job interference and leads to a longer average message latency.
Figure \ref{fig:ur}(a) shows that the 4KB message size has an average message latency of 2.79\textmu s, 16.10\textmu s, and 51.36\textmu s respectively for 3\textmu s, 1\textmu s, and 0.5\textmu s message intervals. The overloaded (0.5\textmu s interval) target application slows 18.4x average message latency compared with the underutilized case (3\textmu s interval). 
The intra-job interference slowdown phenomenon is less severe for the target application with larger message size. The 4MB message results in Figure \ref{fig:ur}(c) indicates the baseline average message latency is of 1.47ms, 2.60ms and 3.62ms respectively for 4ms, 1ms, and 0.7ms message intervals. The maximum slowdown of average message latency is plunged to 2.46x, which is much smaller than the 18.4x slowdown found for 4KB message.
The intra-job interference effect is independent of job placement policies for the target application with UR pattern. Figure \ref{fig:ur}(d) shows that the 4KB message under the random placement has a baseline average message latency of 2.32\textmu s, 2.95\textmu s, and 26.93\textmu s for 3\textmu s, 1\textmu s, and 0.5\textmu s message intervals respectively. The maximum slowdown between the overloaded (0.5\textmu s interval) and underutilized (3\textmu s interval) communication intensities is 11.6x. For 4MB message in Figure \ref{fig:ur}(f), baseline average latency is 1.48ms, 2.84ms, and 3.79ms for 4ms, 1ms, and 0.7ms interval with a maximum of 2.56x slowdown, which is similar to the 2.46x slowdown found under contiguous placement. 

\textbf{\textit{Inter-job interference.}} 
The background application causes the inter-job interference and results in a higher message latency (e.g., the blue, green, and orange boxes shown in the figure).
Results in Figure \ref{fig:ur} shows that inter-job interference can be mitigated with contiguous job placement.

Under the contiguous placement, Figure \ref{fig:ur}(c) shows that the 4MB message with 1ms interval has the average message latency of 2.603ms, 2.702ms, 2.638ms, and 2.793ms for the baseline, the 50\%, 90\%, and 130\% background application respectively. The $130\%$ background application can lead to 1.07x average message latency slowdown, which is the maximum we can observe for the target application under the contiguous placement. The 4KB message and the 512KB message experience similar slowdown, but much less visible.

The random job placement makes inter-job interference more severe, especially for the target application with non-overloaded intensities.
Always compared with the baseline, the 130\% background application causes a longer message latency:
The average message latency of 4KB message with 1\textmu s interval is increased from 2.95\textmu s to 9.33\textmu s which is a 3.16x slowdown in Figure \ref{fig:ur}(d); 
the average message latency of the 512KB message with 150\textmu s interval is surged from 0.13ms to 0.66ms with 5.08x slowdown in Figure \ref{fig:ur}(e);
for 4MB message with 4ms interval, the average message latency is changed from 1.48ms to 2.95ms with 1.99x slowdown in Figure \ref{fig:ur}(f).
Unlike contiguous placement, where we find the 130\% background application causing no more than 7\% target application slowdown, random placement makes the performance variation more obvious. 
This is because under the contiguous placement, the target application with UR pattern is isolated from the background application without any group overlapped. 
Therefore, the majority of the background application's messages does not go across the target application groups. However under random placement, because of group sharing between the target application and the background application, messages of different applications can have the same source or destination group, leading to a network bandwidth competition between them and resulting in a more severe inter-job interference. 
However, Figure \ref{fig:ur}(f) shows that the $130\%$ background application only causes $3.6\%$ slowdown for the 4MB message with 0.7ms interval. This indicates that when the target application message size is large and has an overloaded communication intensity, it can be less affected by other jobs.

In summary, for the target application with UR pattern, if it has an exclusive access to the system, intra-job interference is the main issue impacting the application communication performance.
When running the target application with background traffic, inter-job interference can have a great impact on the application, especially when the target application shares groups with the background application with the random placement. 
Job placement is critical for mitigating inter-job interference. Our results indicate that isolating the target application from the background application with the contiguous placement can reduce the inter-job interference and result in a much less performance degradation.

\subsection{3D Stencil}
Experiment results for the target application with 3D stencil pattern are shown in Figure \ref{fig:stencil}.

\textbf{\textit{Baseline performance.}} 
The performance of 3D stencil pattern is greatly affected by job placement policies. 
Baseline performance shows that the random placement slows the average message latency at the maximum of 22.60x, 10.20x, and 1.54x compared with the contiguous placement for the 4KB, 512KB, and 4MB message respectively.
This is because in 3D stencil pattern, each MPI process communicates with its 6 neighbors on a 3D grid. 
Under the contiguous placement, most of the MPI processes and its 6 neighbors are placed in the same group, thus a large portion of the communication is within group, which saves message inter-group traveling time.
The MPI processes located on the boundary of a group have to communicate with its neighbors sited in other groups, going through one or two global links depending on the adaptive routing decision, which results in a higher message latency.
The random placement distributes MPI processes across the whole system, makes the majority of the messages' source and destination spread between different groups, thus prolongs the average message latency significantly.

\textbf{\textit{Intra-job interference.}} 
Each plot in Figure \ref{fig:stencil} shows that with the same message size, decreasing message interval will increase application communication intensity, and cause more intra-job interference with the result of a higher average message latency. This result is coherent with previous findings for the target application with UR pattern. 

\textbf{\textit{Inter-job interference.}}
Inter-job interference for 3D stencil application can be mitigated with contiguous placement through job isolation.
Figure \ref{fig:stencil}(c) shows that the 130\% background application causes the maximum of $3.20\%$ slowdown for the 4MB message at 1ms interval. 
The results from Figure \ref{fig:stencil}(a) and (b) show that the 4KB and 512KB message experience less than $1\%$ average message slowdown.
Conversely, the random placement makes the target application be more affected by the background application.
Figure \ref{fig:stencil}(d) shows that the average message latency for 4KB message with 3\textmu s interval is 0.23ms for baseline, which is increased to 0.33ms under the 130\% background application, resulting in a 1.43x average message latency slowdown.
Figure \ref{fig:stencil}(e) and (f) show that the 512KB message suffers a maximum of 1.49x average message latency slowdown at 450\textmu s interval and 4MB message suffers a maximum of 1.15x average message latency slowdown at 4ms message interval.

Contiguous placement can mitigate inter-job interference for applications with 3D stencil pattern by minimizing the number of inter-group messages.
In contrast, random placement makes most of the messages of the target application transmitted between groups, and compete for the shared global links with the background application.
Random placement makes the target application more sensitive to communication traffic from other jobs sharing the system. Therefore, in order to have a better baseline performance and mitigate the inter-job interference, applications with 3D Stencil pattern prefers the contiguous job placement.

\subsection{Tornado}

Experiment results of the target application with tornado pattern are shown in Figure \ref{fig:tornado} using the same format as previous figures.

\textbf{\textit{Baseline performance.}} 
The baseline performance in Figure \ref{fig:tornado} shows that tornado pattern prefers random placement. The contiguous placement slows the average message latency at the maximum of 17.32x, 7.96x, and 2.77x compared with the random placement for the 4KB, 512KB, and 4MB message respectively. 
As the tornado pattern in this study makes each MPI process has the communication partner with a fix process ID offset, which is equal to the group size, the contiguous placement places the source and destination processes into neighboring groups and intensifies the usage of the global links between them. In contrast, random placement will make source and destination pairs dispersed across the system, directly makes the traffic more balanced on the network and results in a shorter average message latency.

\textbf{\textit{Intra-job interference.}} 
With the decrease of message interval, the intra-job interference is increased and result in a longer average message latency for baselines both under contiguous and random placement.
Figure \ref{fig:tornado}(a) shows that under contiguous placement, the 4KB message has the baseline average message latency of 4.34\textmu s, 53.12\textmu s, and 125.39\textmu s for 3\textmu s, 1\textmu s, and 0.5\textmu s message intervals. 
The random placement gives a better performance. Figure \ref{fig:tornado}(d) shows that the baseline average message latency for the 4KB message case under random placement is of 2.28\textmu s, 3.06\textmu s and 29.97\textmu s.
The random placement introduces less intra-job interference for the tornado application because this placement helps balance the message traffic across the network. 

\textbf{\textit{Inter-job interference.}} 
Although the random placement can reduce intra-job interference, it causes more inter-job interference and makes the target application more susceptible to the background application.
Figure \ref{fig:tornado-interjob} presents the results of the maximum slowdown of the average message latency caused by the inter-job interference under two placement policies for different target application message sizes and intervals.
Random placement always causes a higher performance degradation. 
Under the contiguous placement, the background application introduces on average of 1.2x maximum slowdown compared with an average of 2.4x maximum slowdown under the random placement. 
On extreme case, the random placement can cause up to 8.22x slowdown in the 512KB case with 150\textmu s interval, whereas contiguous placement only results in a maximum of 1.52x slowdown.

The random placement for the application with tornado communication pattern principally gives a better average performance at the cost of making the target application more sensitive to inter-job interference. For worst case scenario, the average message latency can be prolonged by 8 times compared with baseline performance. On the other hand, contiguous placement introduces larger intra-job interference but greatly reduced inter-job interference for the tornado pattern.

\subsection{Broadcasting}

Message latency of the target application with broadcasting pattern is shown in Figure \ref{fig:bc-inter}. 
In addition, we also collect the inter-group message arrival rate at the spine routers for both the target application and the background application.
Figure \ref{fig:bc-inject} presents the inter-group message arrival rate of the target application in broadcasting root group.
The message arrival rate of the background application is more than 50x greater than the broadcasting rate, thus not shown in the figure. 
Only 4KB message results are presented as the 512KB and 4MB messages have a similar behavior except for a longer message latency. 

\textbf{\textit{Baseline performance.}} 
Three baseline results in Figure \ref{fig:bc-inter} show that the average message latencies under different job placement policies are 0.27ms with less than $1\%$ variation. 
This indicates that job placement can hardly affect broadcasting performance when the target application is the only job in the system. 
As the target application is composed of 2304 processes and each group only has 384 compute nodes, most of the MPI processes are placed among different groups. Therefore, large-scale broadcasting pattern blurs the border between contiguous and random placement by making the majority of messages traverse between groups and resulting in a similar performance.

\begin{figure}[htbp]
\centering
\includegraphics[width=1\linewidth]{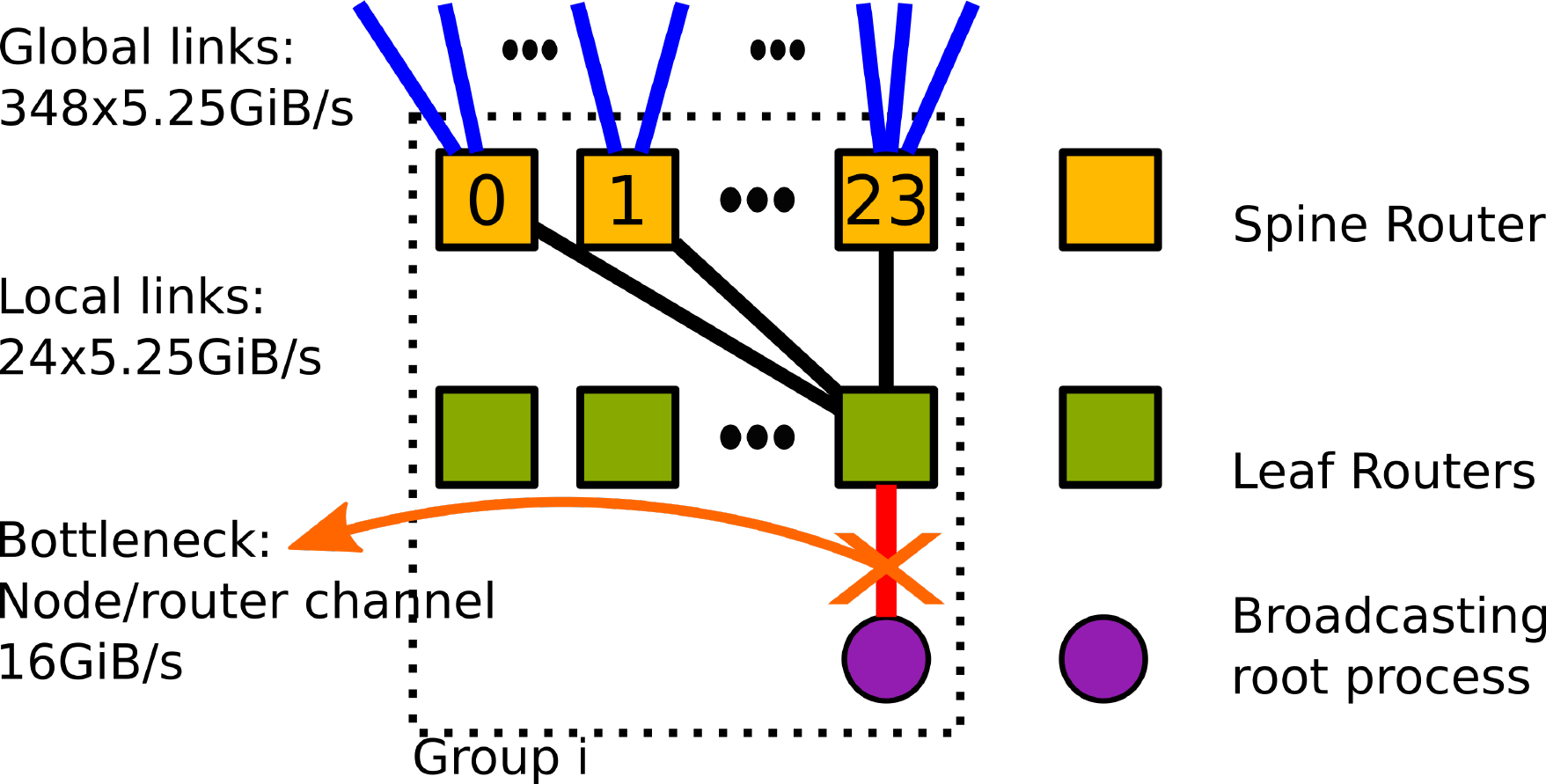}
\caption{The 16GiB/s node-to-router channel is performance bottleneck for the broadcasting pattern.}
\label{fig:bc_limit}
\end{figure}

\textbf{\textit{Intra-job interference.}} 
Intra-job interference is inevitable for broadcasting pattern and the channel between the broadcasting root node and its leaf router is the bottleneck.
As shown in Figure \ref{fig:bc_limit}, the compute node holding the broadcasting root is connected to a leaf router through a single $16GiB/s$ node-to-router channel. As the broadcasting root is the only process generating messages, the target application can only inject traffic to the network at a maximum rate of $16GiB/s$.
However, the leaf router has 24 local links, each of $5.25GiB/s$, to spine routers and the group has 384 global links, each of $4.37GiB/s$ connecting to other groups. Therefore, the target application can never congest local or global links by itself, and the node-to-router channel is the performance bottleneck. All the broadcasting messages compete for this channel, on which the intra-job interference happens.

\textbf{\textit{Inter-job interference.}} 
When there are multiple jobs on the system, the application with broadcasting pattern is more resilient to the background traffic compared with the other patterns. 
The average message latency of the target application under the 520\% background application is 0.27ms, 0.30ms, and 0.47ms for the contiguous placement, and the two random placement cases: the broadcasting root outside and in the background application groups.

Under the contiguous placement, different background application intensities have little influence on the target application with less than $1\%$ average message latency slowdown as shown in Figure \ref{fig:bc-inter}(a).
This phenomenon occurs because under the contiguous placement, the background application cannot congest the network between groups that contain the broadcasting processes.
Therefore, no matter how heavy the background application is, local and global links between the groups holding broadcasting processes are under-utilized. 
The bottleneck of the node-to-router channel makes the broadcasting messages spend most of their transmission time waiting to be sent to the leaf router. 
Once they reach the leaf router, network can send them to their destinations without too much congestion. 
Figure \ref{fig:bc-inject}(a) explains the above observation by listing inter-group message arrival rate of the target application at broadcasting root group under the 520\% background application.
The average broadcasting inter-group message arrival rate is $13.32GiB/s$ under contiguous placement. Notice that because intra-group messages are not recorded, this rate is smaller than $16GiB/s$. 
The 520\% background application message arrival rate is $874.29GiB/s$, which is much smaller than the group's total global link capacity of 384$\times$$4.37GiB/s$.

Under the random placement, the 520\% background application introduces a $11.1\%$ and a $74.1\%$ average message latency slowdown for the cases that the broadcasting root is outside and in background application groups respectively.
In the case that the broadcasting root shares group with the background application, the 520\% background application has an average message arrival rate of $1758.43GiB/s$, which is greater than the group's total global link capacity. This leads to the congestion at the broadcasting root group and slow down its message arrival rate, which is $7.58GiB/s$ on average in Figure \ref{fig:bc-inject}(c). Therefore, comparing the two random placements, sharing group between broadcasting root with the background application can cause an additional 6.67x slowdown.

\section{Conclusion} \label{sec:conclude}

Dragonfly+ is considered as a promising interconnect topology for next-generation supercomputers. 
Although Dragonfly+ networks offer more path diversity than the original Dragonfly design, they are still prone to the performance variability problem due to their hierarchical architecture and resource sharing design \cite{kang2021q, kang2022study, kang2023workload, kang2022workload}. 
In this study, we have enhanced the CODES Dragonfly+ module and have quantitatively evaluated a variety of application communication interactions on a 3,456-node Dragonfly+ system. 
Our study focused on the communication interference from a user's perspective by examining how a target application behaves (i.e., variation of the target application's message latency) under various background application intensities. 


Through this study, we have the following key findings: 

(1) Intra-job interference could be reduced by using different job placement policies depending on the communication pattern of the application.
3D stencil pattern could benefit from contiguous placement by minimizing the number of inter-group transmitted messages. 
Tornado pattern could use random placement to balance the traffics across the network.
For broadcasting pattern, the node-to-router channel is the performance bottleneck where intra-job interference occurs. As such, job placement policies make little performance impact on large-scale broadcasting application,

(2) Inter-job interference problem can generally be mitigated for applications with one-to-one and one-to-many/all communication patterns through job isolation.
The average message latency variation is small for applications with uniform random, 3D stencil and tornado pattern, and applications with large scale broadcasting pattern can hardly be affected by other jobs when they do not share groups with other jobs. 
Moreover, large scale broadcasting is also resilient to the inter-job interference under random placement, even if the broadcasting root process shares groups with other jobs.
\begin{acks}
This work is supported in part by US National Science Foundation grants CNS-1717763, CCF-1422009, CCF-1618776.
This material is based upon work supported by the U.S. Department of Energy, Office of Science, Office of Advanced Scientific Computing Research, under contract number DE-AC02-06CH11357.
\end{acks}

%
\bibliographystyle{ACM-Reference-Format}


\bibliography{sample-base}

%



\end{document}